\documentclass{WileyMSP-template}
\usepackage{amssymb}
\usepackage{cite}
\usepackage{xcolor}
\usepackage{ragged2e}

\begin{document}
\pagestyle{empty}


\title{Chiral Pseudogap Metal Emerging from a Disordered van der Waals Mott Insulator 1$T$-TaS$_{2-x}$Se$_{x}$}

\maketitle


\author{Hyunjin Jung}
\author{Jiwon Jung}
\author{ChoongJae Won}
\author{Hae-Ryong Park}
\author{Sang-Wook Cheong}
\author{Jaeyoung Kim}
\author{Gil Young Cho}
\author{Han Woong Yeom*}



\begin{affiliations}
Mr. Hyunjin Jung, Dr. Jiwon Jung, Dr. ChoongJae Won, Mr. Hae-Ryong Park, Dr. Jaeyoung Kim, Prof. Gil Young Cho, Prof. Han Woong Yeom\\
Center for Artificial Low Dimensional Electronic Systems, Institute for Basic Science (IBS), Pohang 37673, Republic of Korea\\
Email Address: yeom@postech.ac.kr

Mr. Hyunjin Jung, Mr. Hae-Ryong Park, Prof. Gil Young Cho, Prof. Han Woong Yeom\\
Department of Physics, Pohang University of Science and Technology, Pohang 37673, Republic of Korea

Dr. ChoongJae Won, Prof. Sang-Wook Cheong\\
Laboratory for Pohang Emergent Materials, POSTECH, Pohang 37673, Republic of Korea

Dr. ChoongJae Won, Prof. Sang-Wook Cheong\\
MPPC-CPM, Max Planck POSTECH/Korea Research Initiative, Pohang 37673, Republic of Korea

Prof. Sang-Wook Cheong\\
Rutgers Center for emergent Materials and Department of Physics and Astronomy, Rutgers University, NJ, USA

Prof. Gil Young Cho\\
Department of Physics, Korea Advanced Institute of Science and Technology, Daejeon 34141, Republic of Korea 
\end{affiliations}



\begin{abstract}

The emergence of a pseudogap is a hallmark of anomalous electronic states formed through substantial manybody interaction but the mechanism of the pseudogap formation and its role in related emerging quantum states such as unconventional superconductivity remain largely elusive. 
Here, we report the emergence of an unusual pseudogap in a representative van der Waals chiral charge density wave (CDW) materials with strong electron correlation, 1$T$-TaS$_2$, through isoelectronic substitute of S. 
We investigate systematically the evolution of electronic band dispersions of 1$T$-TaS$_{2-x}$Se$_{x}$ (0 $\leqslant$ x  $\leqslant$ 2) using angle-resolved photoemission spectroscopy (ARPES). 
Our results show that the Se substitution induces a quantum transition from an insulating to a pseudogap metallic phase with the CDW order preserved. 
Moreover, the asymmetry of the pseudogap spectral function is found, which reflects the chiral nature of CDW structure.
The present observation is contrasted with the previous suggestions of a Mott transition driven by band width control or charge transfer. 
Instead, we attribute the pseudogap phase to a disordered Mott insulator in line with the recent observation of substantial lateral electronic disorder. 
These findings provide a unique electronic system with chiral pseudogap, where the complex interplay between CDW, chirality, disorder, and electronic correlation may lead to unconventional emergent physics.

\end{abstract}


\section{Introduction}
Pseudogap, suppressed electronic density of states around the Fermi energy, indicates anomalous electronic states departing substantially from the Fermi liquid framework. Pseudogap states have been observed in various materials with strong electronic correlation \cite{kim2011radial, ryu2021pseudogap, battisti2017universality, lahoud2014emergence, gao2020pseudogap, lee2021metal, wang2018disorder, peng2022electronic, borisenko2008pseudogap, chen2017emergence} including a notable example of high-temperature superconductors \cite{timusk1999pseudogap, vershinin2004local, tanaka2006distinct, he2014fermi, hashimoto2015direct, kordyuk2015pseudogap, hanaguri2004checkerboard, kohsaka2007intrinsic, kohsaka2012visualization, loret2019intimate, sacepe2011localization, bastiaans2021direct, tromp2023puddle, zhang2012nodal, kasahara2010evolution, qiu2012robust}. 
A few different mechanisms have been proposed as origins of pseudogap states, which involve competing orders \cite{loret2019intimate, peng2022electronic, borisenko2008pseudogap, chen2017emergence}, preformed Cooper pairs \cite{sacepe2011localization, bastiaans2021direct, tromp2023puddle}, and disorder \cite{kim2011radial, ryu2021pseudogap, gao2020pseudogap, lahoud2014emergence, lee2021metal, sacepe2011localization, bastiaans2021direct, hanaguri2004checkerboard, kohsaka2007intrinsic, kohsaka2012visualization, wang2018disorder, battisti2017universality, zhang2012nodal, kasahara2010evolution, qiu2012robust}. 
Pseudogap states have also been observed in charge-density-wave (CDW) materials with emerging superconductivity such as 1$T$-TaS$_{2}$ \cite{lahoud2014emergence} and 2$H$-TaSe$_{2}$ \cite{borisenko2008pseudogap}, which have been related to the impurity-induced disorder and the CDW fluctuation, respectively. 
The case of 1$T$-TaS$_{2}$ is particularly interesting since this system has been regarded as a Mott insulator in its CDW state and the emerging pseduogap from a Mott insulator is generically related to those in high-temperature superconductors. 
In addition, the strong spin frustration observed in this system \cite{klanjvsek2017high} may provide further connection to psuedogaps in unconventional superconductors. 

Pristine 1$T$-TaS$_{2}$ exhibits a unique CDW structure in a wide temperature range, whose unit cell has 13 Ta atoms of a David-star (DS) structure forming a $(\sqrt{13}\times\sqrt{13})R13.9^{\circ}$ superstructure as depicted in Fig. 1(c). 
The outer six and inner six Ta atoms pair up, leaving an unpaired $d_z$-like electron with local spin moment at the center of the DS cluster. 
The DS distortion is rotated by 13.9 degree from the pristine structure, which break the mirror symmetry to induce chirality, ferrorotational symmetry in more accurate definition, in the CDW state \cite{cheong2018broken, luo2021ultrafast}.
The large separation between neighboring DS clusters significantly suppresses the hopping of central $d_z$ electrons to result in a flat band, which ultimately leads to a Mott gap at least in the 2D limit \cite{imada1998metal}. 
The CDW-Mott phase undergoes a transition to a superconducting phase under pressure \cite{sipos2008mott,dong2021structural} or various chemical dopings \cite{yu2015gate, li2012fe, ang2012real, liu2013superconductivity, liu2014coexistence, ang2015atomistic}.
The mechanism of the emerging superconductivity, however, is not fully understood. 

Moreover, the isoelectronic substitution of S by Se was also found to induce superconductivity, manifests the intriguing nature of the emerging superconductivity in this system. 
The previous angle-resolved photoemission spectroscopy (ARPES) study for varying Se concentration observed the formation of a flat metallic band at the Fermi level and interpreted it as a result of Mott-to-metal transition induced by the increase in band width due to larger Se orbitals \cite{ang2013superconductivity}.
On the other hand, more recent studies with scanning tunneling spectroscopy and density functional theory calculations revealed the existence of substantial unitcell-by-unitcell electronic inhomogeneity \cite{qiao2017mottness,gao2020pseudogap}.
The interpretation of the global electronic state was however different within these studies as an evolution to a normal metal via a charge transfer insulator \cite{qiao2017mottness} or a formation of a correlated metal due to disorder \cite{gao2020pseudogap}. 
In the latter case, the electronic inhomogeneity may act as a source of random potential for each DS cluster to induce a V-shaped pseudogap [Figs. 1(a) and 1(b)] \cite{lahoud2014emergence, battisti2017universality}.
These observations and interpretations apparently indicate distinct physics and prevent the understanding of the emerging metallic phase and the superconductivity from it. 

In the present study, we reinvestigate carefully the evolution of the electronic structure of 1$T$-TaS$_{2-x}$Se$_{x}$ using ARPES to pin down the origin of the metallic state. 
Throughout the entire range of x from 0 to 2, the CDW order is well preserved as shown by  electron diffraction and Ta 4$f$ photoemission spectra. 
In contrast, the Mott band within the CDW gap gradually disappears and, instead, a pseudogap with a power-law decay of the density of states within about 50 meV from the Fermi level.
The pseudo metallic spectral weight is maximized at x=1, coninciding with the highest superconducting temperature, and gradually replaced by the metallic bands of 1$T$-TaSe$_{2}$. 
In addition, the asymmetry of the pseudogap spectral function in momentum space was observed, which can be related to the chirality of the CDW structure. 
This work is consistent with the scenario of a disorder-induced pseudogap metal \cite{gao2020pseudogap} and suggests the possibility of an unconventional superconductivity from a correlated chiral metallic state.

\section{Results and Discussion}
\subsection{Preserved CDW structure in 1$T$-TaS$_{2-x}$Se$_{x}$}
We measured ARPES spectra of valence bands and core-level photoemission spectra for eight different samples with different Se concentrations of x = 0, 0.5, 0.7, 0.9, 1.0, 1.3, 1.7, and 2.0. 
The CDW superstructure is rather well preserved for all Se concentrations, as manifested by the LEED pattern for a maximally disordered case of x=1 shown in Figs. 1(e) and S1. 
A noticeable change in the LEED patterns is that two chiral domains are found in comparable populations while one of them is dominant in the pristine sample. 
This can be explained by the occurrence of misoriented domains connected by domain walls as reported previously after the Se substitution  \cite{ang2015atomistic,gao2020pseudogap}. 
On the other hand, Ta 4$f$ spectra have been well established to exhibit a CDW-induced splitting as shown in Fig. 1(f) \cite{hughes1976charge, hughes1995site, hughes1996lineshapes}, which is basically due to the valence charge redistribution among twelve Ta atoms of a DS CDW unitcell.  
The size of the CDW splitting is slightly reduced by 40 to 60 meV (roughly 5 to 8 $\%$ of the splitting in the pristine sample, Supplementary Fig. S2), for 1$T$-TaSSe. 
This reduced core-level splitting is similar to that for the NCCDW phase of the pristine sample. 
The above results indicates that the local CDW amplitude are only marginally affected by the Se substitution. 
As can be seen in the Supplementary Fig.S3, the Ta 4$f$ core-level photoemission spectra exhibit relatively consistent linewidths across different sample compositions, except for x=0.5, which shows a noticeably larger width. This suggests that the CDW order is well-preserved across different compositions, with x=0.5 being an exception due plausibly to structural disorder onto the CDW order itself in this particular sample.
Note also that the main Ta 4$f$ peak shifts to a higher energy by $\sim$0.1 eV in 1$T$-TaSSe.
This shift is proportional to the Se concentration (see the Supplementary Fig. S2), which can be due to the progressive change of the band structure and the chemical shift from 1$T$-TaS$_2$ to 1$T$-TaSe$_2$. 
The band structure change will be detailed below.

\subsection{Identifying Pseudogap in 1$T$-TaS$_{2-x}$Se$_{x}$}
The evolution of electronic band dispersions around the CDW gap upon the increase of the Se concentration is given in Fig. 2(a). 
The valence band gaped by the CDW formation appears as a M-shaped band with its top at about $\sim$350 meV below the Fermi level in the pristine sample. 
The strong asymmetry due to the chirality can be noticed in the largely different spectral weight for $k$ and $-k$ \cite{yang2022visualization}. 
Within the CDW gap, one can notice two bands with much smaller dispersions, which originate from the surface and bulk components of the so-called lower Hubbard band (LHB) of the Mott insulating phase, with a higher and a lower binding energy, respectively \cite{jung2022surface}. 
This band also exhibits a strongly chiral ARPES intensity distribution \cite{yang2022visualization}. Note that the energy gap between this band and its partner above the Fermi level can be due either to a Mott gap in a monolayer termination or a bonding-antibonding gap in the bilayer termination at the surface \cite{jung2022surface}.
We call these bands as LHBs for convenience, which will be rationalized below. 
As can be seen in the ARPES map of Fig. 2(a) and the energy-distribution curves (EDCs) (cut2) in Fig. 2(c), the valence band maximum moves toward the Fermi level as the Se concentration increases to reach the reported value of the CDW phase of 1T-TaSe$_2$ (see the arrows in Fig. 2(c)) \cite{tian2024dimensionality}.
This is interpreted as a trivially expected change of the band structure due to the compositional change. 

However, unexpected spectral changes are observed for the LHB within the CDW gap.  
One can find that the LHB bands decrease in intensity and broaden in width upon the increase of the Se concentration until it is completely replaced by a uniform spectral feature without clear dispersion covering the whole CDW gap at x = 0.9.
The EDCs shown in Figs. 2(b)-(c) and Supplementary Figs. S4(a)-(b) show that this spectral feature is metallic, reaching to the Fermi level. 
Beyond x = 1.3, this spectral feature disappears as replaced by the well known metallic band of the bulk 1$T$-TaSe$_2$ \cite{tian2024dimensionality, ang2013superconductivity}. 
The symmetrized EDCs in Figs. 3(a)-(b) and Supplementary Figs. S4(c)-(d)) clearly indicate that the metallic spectral feature has substantially reduced spectral weight near the Fermi energy. 
This spectral behavior is not expected for a Fermi liquid but apparently corresponds to a pseudogap with a size of 100 meV.
Note further that there exists momentum asymmetry of the pseudogap spectral weight as clearly indicated in Fig. 2(d).
A similar asymmetry is also noticed in the valence band, which is weaker and in an opposite direction compared to the chiral ARPES distribution  observed in 1$T$-TaS$_{2}$ (Fig. 2(a) and Supplementary Figs. S5 and S6) \cite{yang2022visualization}.
Thus, this observation tells that the populations of the left and right chiral domains has changed from that of the pristine sample as noticed in the LEED patterns (Fig. 1(e) and Figs. 3(d)-(e)). As mentioned above, 1$T$-TaS$_{2}$ exhibits only one chiral domain but 1$T$-TaSSe has two domains with similar intensities but one slightly more populated than the other (Supplementary Fig. S1).
We emphasize that a full 2D momentum maps of the photoelectron intensities in a large energy window were obtained, as partially presented in Fig. 3(f) and Supplementary Figs. S5 and S6. These data tell consistently that the Fermi surface maps (ARPES intensity maps within the CDW band gap) of both 1$T$-TaS$_{2}$ and 1$T$-TaSSe exhibit broken mirror symmetry along both $k_{x}$ and $k_{y}$.    
Then, the asymmetry of pseudogap spectral weight can reasonably be attributed to the chiral wave function of the electrons within the band gap. 
This is natural since all wave functions must reflect the structure chirality of DS CDW structure. 
The strong chiral spectral distribution of the LHB state, which is not an isolated $d_z$ orbital but a cluster orbital, is clearly observed in the pristine sample (see the Fig. 3(f) and Figs. S5-6) \cite{yang2022visualization}. 
Note, however, that there is still some uncertainty as to whether this asymmetry is a definitive signature of true chiral symmetry or ferrorotational symmetry \cite{yang2022visualization, liu2023electrical}.

\subsection{Possible Origins of Pseudogap in 1$T$-TaS$_{2-x}$Se$_{x}$}
We note that the present observation clearly denies the continuous reduction of the Mott gap and its evolution to a charge-transfer gap (LHB moving into the valence band) as suggested in a recent theoretical calculation \cite{qiao2017mottness}.  
At the same time, a normal metallic state with a strong DOS at the Fermi energy is ruled out, which was predicted at x = 1 in that work \cite{qiao2017mottness}. 
To reveal the origin of the pseudogap metallic state further, we conducted the polarization-dependent ARPES measurements for 1$T$-TaSSe with a well developed pseudogap.  
The valence bands of 1$T$-TaS$_{2}$ (or 1T-TaSe$_{2}$) near the Fermi level originate mostly from Ta $5d_{z^{2}}$ orbital, which has distinct polarization selection rules in photoemissionn from S 3$p$  (or Se 4$p$) bands \cite{chen2020strong, yang2022visualization, tian2024dimensionality}.
The polarization dependence of the pseudogap spectral feature follows unambiguously that of the Ta $5d_{z^{2}}$ orbital (Supplementary Figs. S7 and S8) without any substantial contribution from the Se 4$p$ orbital. 
This observation clearly excludes the possibility of the band-width control of the LHB by the $p-d$ orbital hybridization suggested in the previous ARPES work \cite{ang2013superconductivity}. 
In contrast, the size of the CDW gap decreases with increasing Se concentration [Fig. 2(c)], which can be attributed to the enhanced hybridization between Ta $5d$ and Se $4p$ orbitals from that between Ta $5d$ and S $3p$ orbitals at higher binding energies than the CDW gap. 
The anticorrelation between the Se concentration and CDW gap size through $p-d$ orbital hybridization was discussed previously \cite{reshak2005full, ang2013superconductivity, sayers2023exploring}.
The previous ARPES work further claimed the formation of a flat metallic band just below the Fermi energy \cite{ang2013superconductivity}. 
However, the second derivative of the present pseudogap spectral function gives a very similar flat spectral feature below the Fermi level when the second-derivative data processing is applied as in the previous work (Supplementary Fig. S9). 
We thus believe that the flat band in the previous work represents the edge of the pseudogap where the EDCs change their slope abruptly.

Our results also demonstrate that the emerging pseudogap state is not directly related to the reduced CDW order. 
The CDW gap is well defined for the whole range of the Se concentration and its energy scale (200-350 meV from the Fermi energy) is well separated from that of the pseudogap (less than 100 meV) (Fig. 3). 
Moreover, the CDW gap monotonically decreases as the Se concentration increases while the pseudogap appears only around x = 1. 
Furthermore, as shown in Supplementary Figs. S10 and S11, the pseudogap and CDW gap has distinct temperature dependence. 
In pristine 1$T$-TaS$_{2}$, the LHB transits into a dispersive metallic band at 300 K (in the NCCDW phase) and disappears around 370 K (ICCDW phase).
In contrast, in 1$T$-TaSSe, the pseudogap feature persists without any signature of a dispersion upto 400 K.

The summary of the electronic phase transitions is given in Fig. 4 as revised by the present observation. 
As the Se concentration increases in 1$T$-TaS$_{2-x}$Se$_{x}$, we observe a distinct evolution in the symmetrized EDCs from a U-shaped hard gap characteristic of the insulating state (x=0 to 0.5) to a V-shaped pseudogap (x=0.7 to 1.3), and eventually to a normal metallic phase (x=1.7 to 2.0), indicating three distinct quantum states and two quantum transitions (Fig. 3).
It is well established that the Mott gap closes to yield a normal metallic phase in  1$T$-TaSe$_{2}$, which explains the stiff increase of the metallic DOS above x = 1.5.
A notable feature of our data is that the zero-energy DOS shows a sharp increase between x of 0.5 and 0.7, reaching a maximum near x = 1, forming a dome-shaped phase boundary (Fig. 4(b)). 

After ruling out the effects of the CDW gap and the bandwidth change, we can naturally connect the formation of the pseudogap to the Mott phase. 
It has been rather well established that the disorder in a Mott insulator ubiquitously induces pseudogap \cite{hanaguri2004checkerboard, kohsaka2007intrinsic, kohsaka2012visualization,wang2018disorder,battisti2017universality,lahoud2014emergence,gao2020pseudogap,qiao2017mottness} through the interplay of disorder and long-range Coulomb interaction \cite{efros1975coulomb,chiesa2008disorder, lahoud2014emergence}.
Indeed, the strong unitcell-by-unitcell fluctuation of the LHB state was clearly observed in scanning tunneling spectroscopy \cite{gao2020pseudogap,qiao2017mottness}.
While the Mott phase in the pristine sample can be destroyed by the strong interlayer coupling, we assume that the interlayer coupling is largely reduced due to the strong in-plane disorder. 
The S-Se compositional and configurational disorder changes the LHB energy unitcell-by-unitcell \cite{gao2020pseudogap,qiao2017mottness}, which would subsequently disturb the interlayer bonding between DS unitcells. 
This can further be corroborated by the suppression of the stacking order observed in 1T-TaSSe by neutron diffraction \cite{philip2024disorder}.

The fingerprint of disorder can be found in the ARPES spectral width.
We fit the EDCs (Figs. 4(a,c) and Supplementary Fig. S12) with three components for the top valence band and the second and the first layer components of the LHB, labeled P1, P2, and P3, respectively. 
The widths of these peaks follow a similar dome-shaped trend, mirroring the changes in the DOS at the Fermi level. 
Since the increase of the spectral width reflects the disorder strength, we conclude a strong correlation between disorder and the pseudogap at the Fermi level. 
We can further note that the spectral changes  of the LHB band exhibit the characteristics of the theoretical prediction for a disordered Mott insulator \cite{lahoud2014emergence, chiesa2008disorder}; the reduction of the coherent Mott peaks, the reduction of the Mott gap, the formation of a V-shape gap, and the increase of the Fermi-level DOS. 
All these observations converge consistently to the picture of a pseudogap induced by disorder. 
Since the present system exhibits emerging superconductivity in the range of the Se concentration where the pseudogap develops, we suggest that the superconductivity is based on a disordered correlated metallic state. 
The sign of the multifractality of the metallic wave functions observed in the previous scanning tunneling spectroscopy study \cite{gao2020pseudogap} is consistent with the enhanced superconductivity in a disordered system \cite{zhao2019disorder, lee2024interplay}.

\subsection{Implications of Pseudogap for Symmetry-Broken Superconductivity}

One unique feature of the present system is that the pseudo-metallic wave function exhibits chirality (see the Fermi surface map in Fig. 3(f) and Figs. S5-6) and thus the emerging superconductivity is naturally expected to inherit the broken symmetry of the pseudogap state, opening up intriguing possibilities. 
At the bottom line, this naturally suppresses the formation of isotropic \textit{s}-wave superconductivity, and thus it calls for investigation of the nature of the superconducting state emerging in 1$T$-TaS$_2$.
Note that the symmetry-broken pseudogap state is an important ingredient of the cuprate superconductor \cite{xia2008polar, sato2017thermodynamic, zhang2018discovery, ishida2020divergent}. While such symmetry breaking in cuprates is a spontaneous one through a quantum phase transition, the present one is imprinted \textit{a priori} in the underlying CDW structure. Whether this chiral pseudogap in our case competes or intertwines with the superconductivity is an interesting theoretical and experimental question. Notably, broken reflection symmetry, when combined with broken inversion and time-reversal symmetry, can exhibit the superconducting diode effect, offering promising applications in electronic and superconducting devices \cite{zinkl2022symmetry, nadeem2023superconducting}. It would be of great interesting to look for a nontrivial consequence of the symmetry-broken in the pseudogap phase and the emerging superconductivity from it.

\section{Conclusion}
We investigated the band structures of 1$T$-TaS$_{2-x}$Se$_{x}$ using ARPES for a whole range of the Se concentration. We observed clearly two quantum phase transitions from a Mott insulator to a pseudogap metal and to a normal metal as the Se concetration increases.  
The pseudogap is not related to the band width change, the charge transfer, or the CDW order but to the disorder in electronic states caused by the random substitution of Se. 
We thus interpret that the pseudogap represents a correlated metallic state emerging in a disordered Mott insulator.
We further observe that the pseudogap spectral function exhibits asymmetry in momentum reflecting the chiral CDW structure. 
Since the pseudogap phase coincides the emerging superconducting phase, we speculate that the emerging superconductivity may exhibit unconventional properties. 
The present system can be an ideal platform to study quantum phases of electrons under the complex interplay of electron correlation, CDW, and chirality with a tunable disorder.


\section{Experimental Section}
ARPES and LEED measurements were conducted at the elliptical-undulator beamline 4A2 of the Pohang Light Source (Pohang, Korea). We utilized a wide-angle electron analyzer (DA-30-L, Scienta-Omicron) and a linearly-polarized photon beam of 65 eV for the primary ARPES experiments. In addition, XPS core-level spectra were acquired at both 65 eV (x=0, 0.7, 0.9, 1.0, 1.3, and 1.7) and 200 eV (x=0.5 and 2.0) photon energies. LEED measurements were performed at 300 K with an electron energy of 70 eV.

Single crystals of the 1$T$-TaS$_{2-x}$Se$_{x}$ series were synthesized via the chemical vapor transport method using polycrystalline TaS$_{2-x}$Se$_{x}$ and I$_{2}$ as the transport agent \cite{cho2016nanoscale}. The samples were cleaved in situ at 300 K under ultra-high vacuum conditions (better than 1$\times$10$^{-10}$ Torr) for temperature-dependent ARPES measurements conducted over a range of 70 to 400 K. The sample thickness after the cleavages was a few tens of micrometers.

For data analysis, the pseudogap was quantified by fitting the symmetrized energy distribution curves (EDCs) near the Fermi level using a power-law function of the form $I(E)=I_{0}|E-E_{F}|^{\alpha}+Background$, where $\alpha$ is the exponent that characterizes the V-shaped gap. Additionally, the band widths were determined by fitting the EDCs with a pseudo-Voigt function.

\medskip
\textbf{Acknowledgements} \par 
This work was supported by the Institute for Basic Science (Grant No. IBS-R014-D1). C.J.W. was supported by the National Research Foundation of Korea funded by the Ministry of Science and ICT (Grant No. 2022M3H4A1A04074153).

\medskip
\textbf{Author Contributions} \par
H.W.Y. conceived the project idea and plan. H.J., J.J., and J.K. performed the ARPES measurements. C.W. synthesized the 1$T$-TaS$_{2-x}$Se$_{x}$ single crystals with help from S.W.C.; H.J., H.P., and H.W.Y. analyzed the data. H.J., G.Y.C. and H.W.Y. wrote the paper.

\medskip
\textbf{Data Availability} \par
The data used for Figs. 1-4 and Figs. S1-12 are fully available on request from the corresponding author.

\medskip
\bibliographystyle{MSP}
\bibliography{PG_TSS}

\providecommand{\noopsort}[1]{}\providecommand{\singleletter}[1]{#1}%
\begin{thebibliography}{10}
\providecommand{\url}[1]{\texttt{#1}}
\providecommand{\urlprefix}{URL }

\bibitem{kim2011radial}
K.~S. Kim, H.~W. Yeom,
\newblock \emph{Physical review letters} \textbf{2011}, \emph{107}, 13 136402.

\bibitem{ryu2021pseudogap}
S.~H. Ryu, M.~Huh, D.~Y. Park, C.~Jozwiak, E.~Rotenberg, A.~Bostwick, K.~S. Kim,
\newblock \emph{Nature} \textbf{2021}, \emph{596}, 7870 68.

\bibitem{battisti2017universality}
I.~Battisti, K.~M. Bastiaans, V.~Fedoseev, A.~De~La~Torre, N.~Iliopoulos, A.~Tamai, E.~C. Hunter, R.~S. Perry, J.~Zaanen, F.~Baumberger, et~al.,
\newblock \emph{Nature physics} \textbf{2017}, \emph{13}, 1 21.

\bibitem{lahoud2014emergence}
E.~Lahoud, O.~N. Meetei, K.~Chaska, A.~Kanigel, N.~Trivedi,
\newblock \emph{Physical review letters} \textbf{2014}, \emph{112}, 20 206402.

\bibitem{gao2020pseudogap}
J.~Gao, J.~W. Park, K.~Kim, S.~K. Song, H.~R. Park, J.~Lee, J.~Park, F.~Chen, X.~Luo, Y.~Sun, et~al.,
\newblock \emph{Nano Letters} \textbf{2020}, \emph{20}, 9 6299.

\bibitem{lee2021metal}
K.~Lee, J.~Choe, D.~Iaia, J.~Li, J.~Zhao, M.~Shi, J.~Ma, M.~Yao, Z.~Wang, C.-L. Huang, et~al.,
\newblock \emph{npj Quantum Materials} \textbf{2021}, \emph{6}, 1 8.

\bibitem{wang2018disorder}
Z.~Wang, Y.~Okada, J.~O’Neal, W.~Zhou, D.~Walkup, C.~Dhital, T.~Hogan, P.~Clancy, Y.-J. Kim, Y.~Hu, et~al.,
\newblock \emph{Proceedings of the National Academy of Sciences} \textbf{2018}, \emph{115}, 44 11198.

\bibitem{peng2022electronic}
S.~Peng, C.~Lane, Y.~Hu, M.~Guo, X.~Chen, Z.~Sun, M.~Hashimoto, D.~Lu, Z.-X. Shen, T.~Wu, et~al.,
\newblock \emph{npj Quantum Materials} \textbf{2022}, \emph{7}, 1 58.

\bibitem{borisenko2008pseudogap}
S.~Borisenko, A.~Kordyuk, A.~Yaresko, V.~Zabolotnyy, D.~Inosov, R.~Schuster, B.~B{\"u}chner, R.~Weber, R.~Follath, L.~Patthey, et~al.,
\newblock \emph{Physical review letters} \textbf{2008}, \emph{100}, 19 196402.

\bibitem{chen2017emergence}
P.~Chen, W.~W. Pai, Y.-H. Chan, A.~Takayama, C.-Z. Xu, A.~Karn, S.~Hasegawa, M.-Y. Chou, S.-K. Mo, A.-V. Fedorov, et~al.,
\newblock \emph{Nature communications} \textbf{2017}, \emph{8}, 1 516.

\bibitem{timusk1999pseudogap}
T.~Timusk, B.~Statt,
\newblock \emph{Reports on Progress in Physics} \textbf{1999}, \emph{62}, 1 61.

\bibitem{vershinin2004local}
M.~Vershinin, S.~Misra, S.~Ono, Y.~Abe, Y.~Ando, A.~Yazdani,
\newblock \emph{Science} \textbf{2004}, \emph{303}, 5666 1995.

\bibitem{tanaka2006distinct}
K.~Tanaka, W.~Lee, D.~Lu, A.~Fujimori, T.~Fujii, Risdiana, I.~Terasaki, D.~Scalapino, T.~Devereaux, Z.~Hussain, et~al.,
\newblock \emph{Science} \textbf{2006}, \emph{314}, 5807 1910.

\bibitem{he2014fermi}
Y.~He, Y.~Yin, M.~Zech, A.~Soumyanarayanan, M.~M. Yee, T.~Williams, M.~Boyer, K.~Chatterjee, W.~Wise, I.~Zeljkovic, et~al.,
\newblock \emph{Science} \textbf{2014}, \emph{344}, 6184 608.

\bibitem{hashimoto2015direct}
M.~Hashimoto, E.~A. Nowadnick, R.-H. He, I.~M. Vishik, B.~Moritz, Y.~He, K.~Tanaka, R.~G. Moore, D.~Lu, Y.~Yoshida, et~al.,
\newblock \emph{Nature materials} \textbf{2015}, \emph{14}, 1 37.

\bibitem{kordyuk2015pseudogap}
A.~Kordyuk,
\newblock \emph{Low Temperature Physics} \textbf{2015}, \emph{41}, 5 319.

\bibitem{hanaguri2004checkerboard}
T.~Hanaguri, C.~Lupien, Y.~Kohsaka, D.-H. Lee, M.~Azuma, M.~Takano, H.~Takagi, J.~C. Davis,
\newblock \emph{Nature} \textbf{2004}, \emph{430}, 7003 1001.

\bibitem{kohsaka2007intrinsic}
Y.~Kohsaka, C.~Taylor, K.~Fujita, A.~Schmidt, C.~Lupien, T.~Hanaguri, M.~Azuma, M.~Takano, H.~Eisaki, H.~Takagi, et~al.,
\newblock \emph{Science} \textbf{2007}, \emph{315}, 5817 1380.

\bibitem{kohsaka2012visualization}
Y.~Kohsaka, T.~Hanaguri, M.~Azuma, M.~Takano, J.~Davis, H.~Takagi,
\newblock \emph{Nature physics} \textbf{2012}, \emph{8}, 7 534.

\bibitem{loret2019intimate}
B.~Loret, N.~Auvray, Y.~Gallais, M.~Cazayous, A.~Forget, D.~Colson, M.-H. Julien, I.~Paul, M.~Civelli, A.~Sacuto,
\newblock \emph{Nature Physics} \textbf{2019}, \emph{15}, 8 771.

\bibitem{sacepe2011localization}
B.~Sac{\'e}p{\'e}, T.~Dubouchet, C.~Chapelier, M.~Sanquer, M.~Ovadia, D.~Shahar, M.~Feigel’Man, L.~Ioffe,
\newblock \emph{Nature Physics} \textbf{2011}, \emph{7}, 3 239.

\bibitem{bastiaans2021direct}
K.~M. Bastiaans, D.~Chatzopoulos, J.-F. Ge, D.~Cho, W.~O. Tromp, J.~M. van Ruitenbeek, M.~H. Fischer, P.~J. de~Visser, D.~J. Thoen, E.~F. Driessen, et~al.,
\newblock \emph{Science} \textbf{2021}, \emph{374}, 6567 608.

\bibitem{tromp2023puddle}
W.~O. Tromp, T.~Benschop, J.-F. Ge, I.~Battisti, K.~M. Bastiaans, D.~Chatzopoulos, A.~H. Vervloet, S.~Smit, E.~van Heumen, M.~S. Golden, et~al.,
\newblock \emph{Nature Materials} \textbf{2023}, \emph{22}, 6 703.

\bibitem{zhang2012nodal}
Y.~Zhang, Z.~Ye, Q.~Ge, F.~Chen, J.~Jiang, M.~Xu, B.~Xie, D.~Feng,
\newblock \emph{Nature Physics} \textbf{2012}, \emph{8}, 5 371.

\bibitem{kasahara2010evolution}
S.~Kasahara, T.~Shibauchi, K.~Hashimoto, K.~Ikada, S.~Tonegawa, R.~Okazaki, H.~Shishido, H.~Ikeda, H.~Takeya, K.~Hirata, et~al.,
\newblock \emph{Physical Review B—Condensed Matter and Materials Physics} \textbf{2010}, \emph{81}, 18 184519.

\bibitem{qiu2012robust}
X.~Qiu, S.~Zhou, H.~Zhang, B.~Pan, X.~Hong, Y.~Dai, M.~J. Eom, J.~S. Kim, Z.~Ye, Y.~Zhang, et~al.,
\newblock \emph{Physical Review X} \textbf{2012}, \emph{2}, 1 011010.

\bibitem{klanjvsek2017high}
M.~Klanj{\v{s}}ek, A.~Zorko, R.~{\v{Z}}itko, J.~Mravlje, Z.~Jagli{\v{c}}i{\'c}, P.~K. Biswas, P.~Prelov{\v{s}}ek, D.~Mihailovic, D.~Ar{\v{c}}on,
\newblock \emph{Nature Physics} \textbf{2017}, \emph{13}, 11 1130.

\bibitem{cheong2018broken}
S.-W. Cheong, D.~Talbayev, V.~Kiryukhin, A.~Saxena,
\newblock \emph{npj Quantum Materials} \textbf{2018}, \emph{3}, 1 19.

\bibitem{luo2021ultrafast}
X.~Luo, D.~Obeysekera, C.~Won, S.~H. Sung, N.~Schnitzer, R.~Hovden, S.-W. Cheong, J.~Yang, K.~Sun, L.~Zhao,
\newblock \emph{Physical review letters} \textbf{2021}, \emph{127}, 12 126401.

\bibitem{imada1998metal}
M.~Imada, A.~Fujimori, Y.~Tokura,
\newblock \emph{Reviews of modern physics} \textbf{1998}, \emph{70}, 4 1039.

\bibitem{sipos2008mott}
B.~Sipos, A.~F. Kusmartseva, A.~Akrap, H.~Berger, L.~Forr{\'o}, E.~Tuti{\v{s}},
\newblock \emph{Nature materials} \textbf{2008}, \emph{7}, 12 960.

\bibitem{dong2021structural}
Q.~Dong, Q.~Li, S.~Li, X.~Shi, S.~Niu, S.~Liu, R.~Liu, B.~Liu, X.~Luo, J.~Si, et~al.,
\newblock \emph{npj Quantum Materials} \textbf{2021}, \emph{6}, 1 20.

\bibitem{yu2015gate}
Y.~Yu, F.~Yang, X.~F. Lu, Y.~J. Yan, Y.-H. Cho, L.~Ma, X.~Niu, S.~Kim, Y.-W. Son, D.~Feng, et~al.,
\newblock \emph{Nature nanotechnology} \textbf{2015}, \emph{10}, 3 270.

\bibitem{li2012fe}
L.~Li, W.~Lu, X.~Zhu, L.~Ling, Z.~Qu, Y.~Sun,
\newblock \emph{Europhysics Letters} \textbf{2012}, \emph{97}, 6 67005.

\bibitem{ang2012real}
R.~Ang, Y.~Tanaka, E.~Ieki, K.~Nakayama, T.~Sato, L.~Li, W.~Lu, Y.~Sun, T.~Takahashi,
\newblock \emph{Physical review letters} \textbf{2012}, \emph{109}, 17 176403.

\bibitem{liu2013superconductivity}
Y.~Liu, R.~Ang, W.~Lu, W.~Song, L.~Li, Y.~Sun,
\newblock \emph{Applied Physics Letters} \textbf{2013}, \emph{102}, 19.

\bibitem{liu2014coexistence}
Y.~Liu, W.~Lu, L.~Li, R.~Ang, Y.~Sun,
\newblock \emph{Applied Physics Letters} \textbf{2014}, \emph{104}, 25.

\bibitem{ang2015atomistic}
R.~Ang, Z.~Wang, C.~Chen, J.~Tang, N.~Liu, Y.~Liu, W.~Lu, Y.~Sun, T.~Mori, Y.~Ikuhara,
\newblock \emph{Nature communications} \textbf{2015}, \emph{6}, 1 6091.

\bibitem{ang2013superconductivity}
R.~Ang, Y.~Miyata, E.~Ieki, K.~Nakayama, T.~Sato, Y.~Liu, W.~Lu, Y.~Sun, T.~Takahashi,
\newblock \emph{Physical Review B} \textbf{2013}, \emph{88}, 11 115145.

\bibitem{qiao2017mottness}
S.~Qiao, X.~Li, N.~Wang, W.~Ruan, C.~Ye, P.~Cai, Z.~Hao, H.~Yao, X.~Chen, J.~Wu, et~al.,
\newblock \emph{Physical Review X} \textbf{2017}, \emph{7}, 4 041054.

\bibitem{hughes1976charge}
H.~Hughes, R.~Pollak,
\newblock \emph{Philosophical Magazine} \textbf{1976}, \emph{34}, 6 1025.

\bibitem{hughes1995site}
H.~Hughes, J.~Scarfe,
\newblock \emph{Physical review letters} \textbf{1995}, \emph{74}, 15 3069.

\bibitem{hughes1996lineshapes}
H.~Hughes, J.~Scarfe,
\newblock \emph{Journal of Physics: Condensed Matter} \textbf{1996}, \emph{8}, 10 1457.

\bibitem{yang2022visualization}
H.~Yang, K.~He, J.~Koo, S.~Shen, S.~Zhang, G.~Liu, Y.~Liu, C.~Chen, A.~Liang, K.~Huang, et~al.,
\newblock \emph{Physical review letters} \textbf{2022}, \emph{129}, 15 156401.

\bibitem{jung2022surface}
J.~Jung, J.~W. Park, J.~Kim, H.~W. Yeom,
\newblock \emph{Physical Review B} \textbf{2022}, \emph{106}, 15 155406.

\bibitem{tian2024dimensionality}
N.~Tian, Z.~Huang, B.~G. Jang, S.~Guo, Y.-J. Yan, J.~Gao, Y.~Yu, J.~Hwang, C.~Tang, M.~Wang, et~al.,
\newblock \emph{National Science Review} \textbf{2024}, \emph{11}, 3 nwad144.

\bibitem{liu2023electrical}
G.~Liu, T.~Qiu, K.~He, Y.~Liu, D.~Lin, Z.~Ma, Z.~Huang, W.~Tang, J.~Xu, K.~Watanabe, et~al.,
\newblock \emph{Nature nanotechnology} \textbf{2023}, \emph{18}, 8 854.

\bibitem{chen2020strong}
Y.~Chen, W.~Ruan, M.~Wu, S.~Tang, H.~Ryu, H.-Z. Tsai, R.~L. Lee, S.~Kahn, F.~Liou, C.~Jia, et~al.,
\newblock \emph{Nature Physics} \textbf{2020}, \emph{16}, 2 218.

\bibitem{reshak2005full}
A.~H. Reshak, S.~Auluck,
\newblock \emph{Physica B: Condensed Matter} \textbf{2005}, \emph{358}, 1-4 158.

\bibitem{sayers2023exploring}
C.~Sayers, G.~Cerullo, Y.~Zhang, C.~Sanders, R.~Chapman, A.~Wyatt, G.~Chatterjee, E.~Springate, D.~Wolverson, E.~Da~Como, et~al.,
\newblock \emph{Physical Review Letters} \textbf{2023}, \emph{130}, 15 156401.

\bibitem{efros1975coulomb}
A.~L. {\'E}fros, B.~I. Shklovskii,
\newblock \emph{Journal of Physics C: Solid State Physics} \textbf{1975}, \emph{8}, 4 L49.

\bibitem{chiesa2008disorder}
S.~Chiesa, P.~B. Chakraborty, W.~E. Pickett, R.~T. Scalettar,
\newblock \emph{Physical review letters} \textbf{2008}, \emph{101}, 8 086401.

\bibitem{philip2024disorder}
S.~S. Philip, D.~Louca,
\newblock \emph{Physical Review B} \textbf{2024}, \emph{109}, 9 094118.

\bibitem{zhao2019disorder}
K.~Zhao, H.~Lin, X.~Xiao, W.~Huang, W.~Yao, M.~Yan, Y.~Xing, Q.~Zhang, Z.-X. Li, S.~Hoshino, et~al.,
\newblock \emph{Nature Physics} \textbf{2019}, \emph{15}, 9 904.

\bibitem{lee2024interplay}
J.~Lee, H.-R. Park, J.~S. Kim, H.~W. Yeom,
\newblock \emph{Advanced Materials} \textbf{2024}, \emph{36}, 35 2404708.

\bibitem{xia2008polar}
J.~Xia, E.~Schemm, G.~Deutscher, S.~Kivelson, D.~Bonn, W.~Hardy, R.~Liang, W.~Siemons, G.~Koster, M.~Fejer, et~al.,
\newblock \emph{Physical Review Letters} \textbf{2008}, \emph{100}, 12 127002.

\bibitem{sato2017thermodynamic}
Y.~Sato, S.~Kasahara, H.~Murayama, Y.~Kasahara, E.-G. Moon, T.~Nishizaki, T.~Loew, J.~Porras, B.~Keimer, T.~Shibauchi, et~al.,
\newblock \emph{Nature Physics} \textbf{2017}, \emph{13}, 11 1074.

\bibitem{zhang2018discovery}
J.~Zhang, Z.~Ding, C.~Tan, K.~Huang, O.~O. Bernal, P.-C. Ho, G.~D. Morris, A.~D. Hillier, P.~K. Biswas, S.~P. Cottrell, et~al.,
\newblock \emph{Science advances} \textbf{2018}, \emph{4}, 1 eaao5235.

\bibitem{ishida2020divergent}
K.~Ishida, S.~Hosoi, Y.~Teramoto, T.~Usui, Y.~Mizukami, K.~Itaka, Y.~Matsuda, T.~Watanabe, T.~Shibauchi,
\newblock \emph{Journal of the Physical Society of Japan} \textbf{2020}, \emph{89}, 6 064707.

\bibitem{zinkl2022symmetry}
B.~Zinkl, K.~Hamamoto, M.~Sigrist,
\newblock \emph{Physical Review Research} \textbf{2022}, \emph{4}, 3 033167.

\bibitem{nadeem2023superconducting}
M.~Nadeem, M.~S. Fuhrer, X.~Wang,
\newblock \emph{Nature Reviews Physics} \textbf{2023}, \emph{5}, 10 558.

\bibitem{cho2016nanoscale}
D.~Cho, S.~Cheon, K.-S. Kim, S.-H. Lee, Y.-H. Cho, S.-W. Cheong, H.~W. Yeom,
\newblock \emph{Nature communications} \textbf{2016}, \emph{7}, 1 10453.

\end{thebibliography}

\begin{figure*} [htb!]
\includegraphics{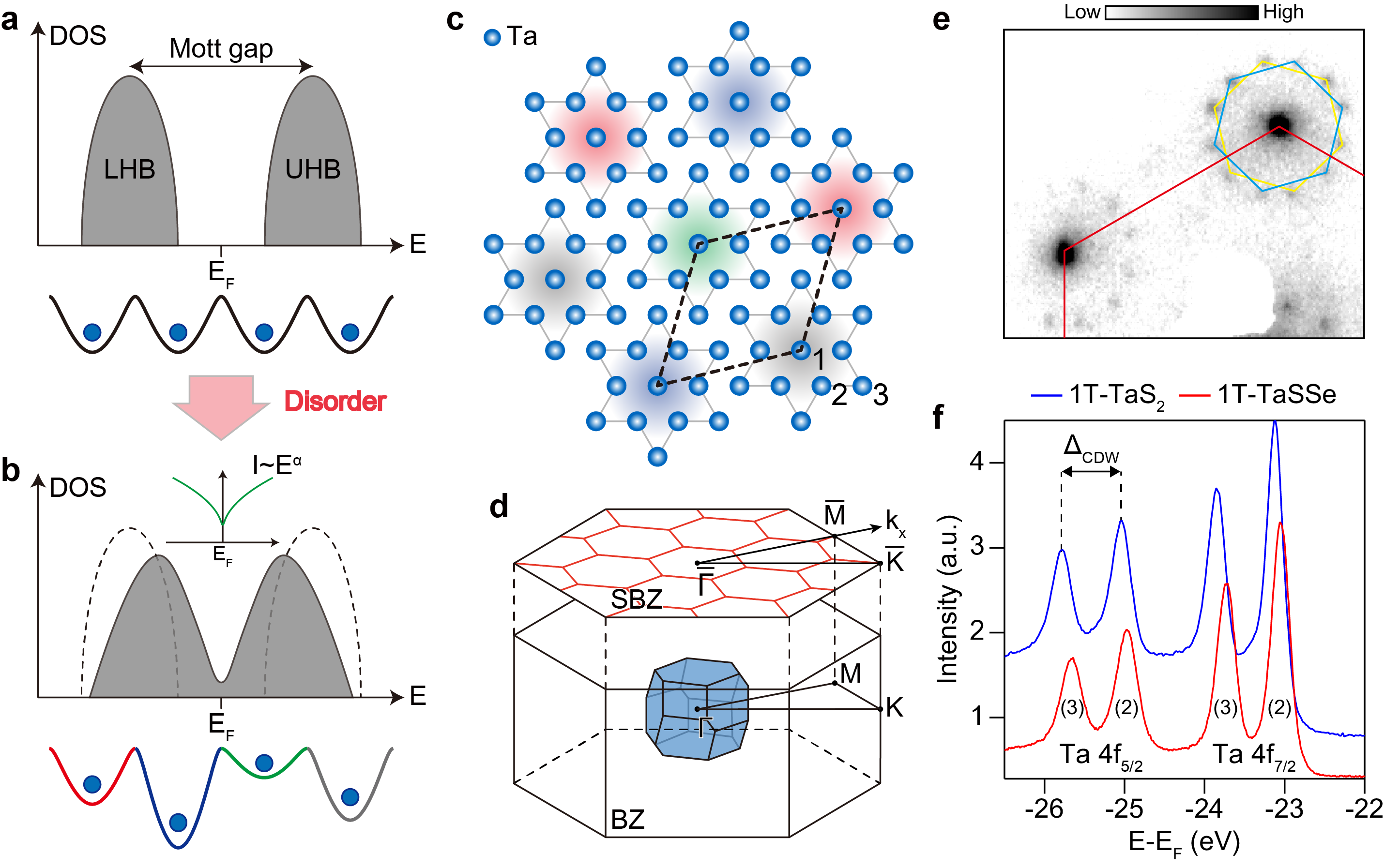}
\centering
\caption{\label{figure 1}
Crystalline structure and CDW phase in 1$T$-TaSSe. (a), Schematic density of states (DOS) in Mott insulator, where electrons experience a homogeneous Coulomb potential, leading to the formation of a Mott gap. (b), Schematic DOS in disordered Mott insulator, where random potential induced by disorder lead to the emergence of a V-shaped gap within the Mott gap. (c), Reconstructed atomic structure of Ta atoms, illustrating disorder effects on the CDW sturcture. (d), Bulk and surface Brillouin zones of 1$T$-TaS$_{2-x}$Se$_{x}$. (e), LEED image of 1$T$-TaSSe with electron energy of 70 eV at 300 K. The yellow and blue guidelines show the diffraction spots generated by two different degenerate CDW domains rotated by 27.8 degrees, matched well with FFT image of STM \cite{gao2020pseudogap}. (f), Direct comparison of Ta 4$f$ core-level spectra between 1$T$-TaS$_{2}$ (blue) and 1$T$-TaSSe (red) at 70 K. The CDW splitting is indicated by $\Delta_{CDW}$, which is 740 meV in 1$T$-TaS$_{2}$, and 680 meV in 1$T$-TaSSe, respectively. The components labeled b and c correspond to the contributions from the inner and outer six Ta atoms in the DS cluster, respectively.}
\end{figure*}

\begin{figure*} [htb!]
\includegraphics{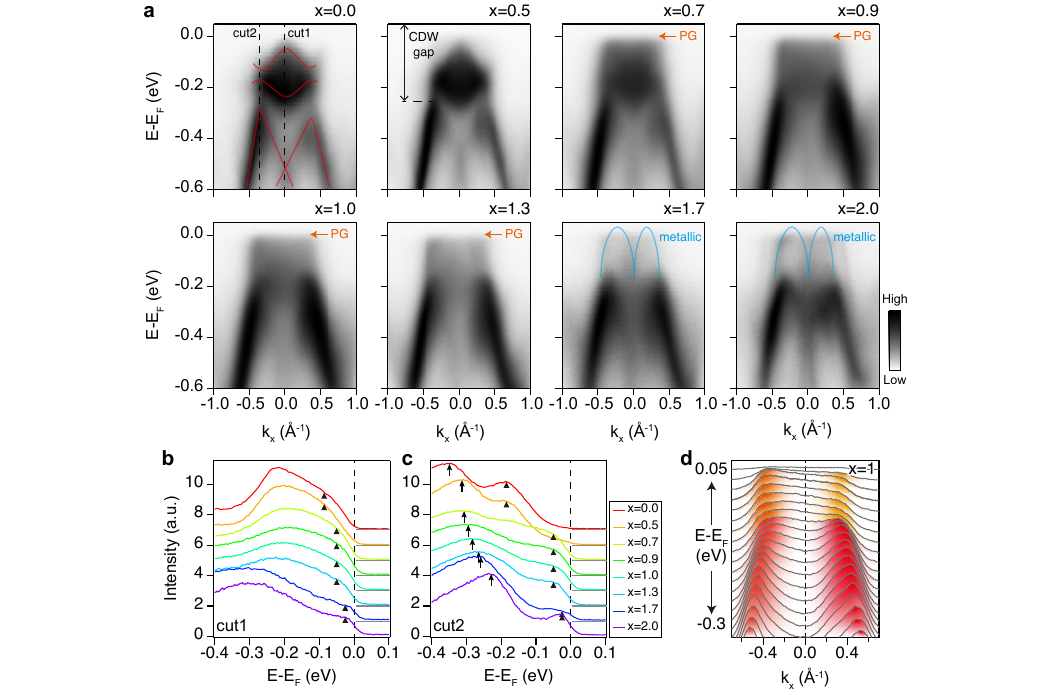}
\centering
\caption{\label{figure 2}
Evolution of electronic band structure of 1$T$-TaS$_{2-x}$Se$_{x}$ along $\overline{\Gamma}$-$\overline{M}$ direction. (a), Experimentally observed band dispersion for various Se concentration (x=0.0, 0.5, 0.7, 1.0, 1.3, 1.7, and 2.0). M-shaped and two flat bands are illustrated by red lines (x=0), pseudogap (PG) are highlighted by the orange arrows (x=0.7-1.3), and the emergence of metallic bands is
described by blue curves (x=1.7-2.0). (b-c), Energy distribution curves (EDCs) corresponding to cut1 and cut2, respectively, with respect to Se concentration. Cut1 intersects the $\Gamma$ point (k$_{x}$=0), whereas cut2 crosses the CDW gap (k$_{x}$=-0.37 Å$^{-1}$). The black triangles denote the leading edges of EDCs near the Fermi level, and the black arrows highlight the CDW gaps. The gray lines indicate the zero-point for each dataset. Each EDC spectrum was normalized by dividing by the sum of the intensity in the range of -0.4 to 0 eV.
(d), Momentum distribution curves (MDCs) at x=1, highlighting the asymmetry in spectral intensity near the Fermi level. The pseudogap contribution near the Fermi level exhibits higher intensity along the $-k$ direction, while the valence band top shows stronger intensity along the $+k$ direction.}
\end{figure*}

\begin{figure*} [htb!]
\includegraphics{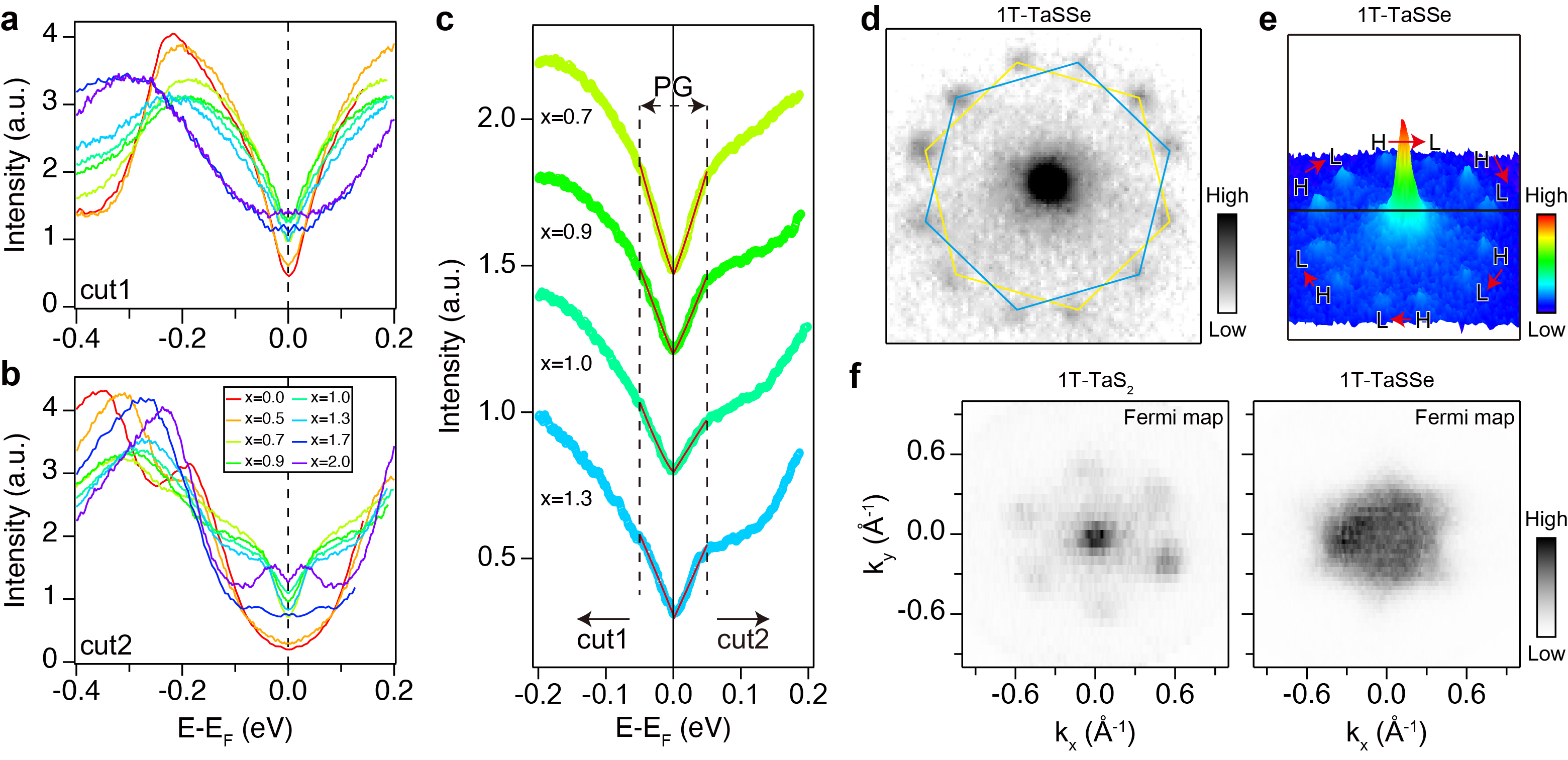}
\centering
\caption{\label{figure 3}
Characterization of pseudogap states in 1$T$-TaS$_{2-x}$Se$_{x}$. 
(a-b), Symmetrized energy distribution curves (EDCs) near the Fermi level for cut1 and cut2, respectively, showing the evolution of spectral intensity with increasing Se concentration.
(c), The zoom-in of symmetrized EDCs in the pseudogap region for Se concentrations of x=0.7, 0.9, 1.0, and 1.3. The red lines represent fitted power-law function, $I(E)=I_{0} \cdot |E-E_{F}|^{\alpha}+BG$, where $I_{0}$ corresponds to the intensity and $BG$ represents the background. The mean value of the fitted power-law exponents is determined to be $\alpha = 1.0708 \pm 0.0317$.
(d), Zoom-in of the LEED patterns in Fig. 1(e). Yellow and blue hexagonal lines show two different David-star clusters.
(e), Intensity mapping of (d), illustrating intensity variations in domain populations. High population domain (H) and low population domain (L) are marked.
(f), Constant energy contour maps of 1$T$-TaS$_{2}$ (left) and 1$T$-TaSSe (right) at the Fermi level. The ARPES maps at different binding energies are provided in the supplementary Fig. S5.
}
\end{figure*}

\begin{figure*} [t!]
\includegraphics{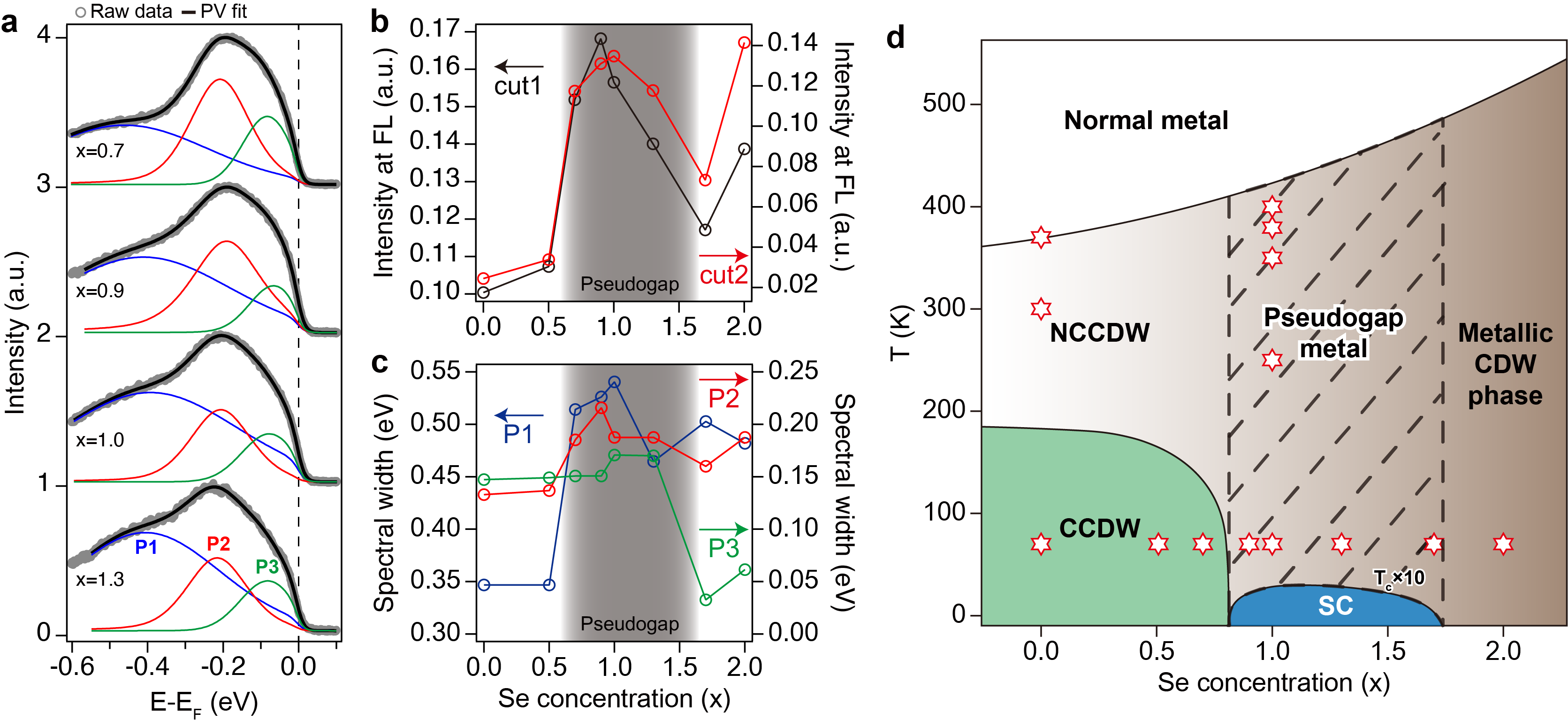}
\centering
\caption{\label{figure 4}
Revised phase diagram of 1$T$-TaS$_{2-x}$Se$_{x}$. 
(a), Pseudo-Voight (PG) fits of EDC spectra at $\overline{\Gamma}$ point (cut1) in the pseudogap region for x=0.7, 0.9, 1.0, and 1.3. The full evolution of PG fits are in the Supplementary Fig. S12.
(b), ARPES spectral intensity for cut1 (black) and cut2 (red) at the Fermi level as a function of Se concentration, highlighting the pseudogap region. 
The intensity at Fermi level was obtained by integrating the symmetrized EDCs over the energy range of -50 to +50 meV. 
(c), Evolution of FWHM for P1, P2, and P3, representing the spectral width, as a function of Se concentration. The spectra were analyzed by fitting three distinct peaks (P1, P2, P3) using pseudo-Voigt function (see details in Fig. 4(a) and Fig. S8).
(d), The diagram illustrates different phases, including CCDW, NCCDW, pseudogap metal, normal metal, and superconducting (SC) regions, with the highest $T_{c}$ observed at x=1. The red David-stars represent experimentally measured data points plotted on the phase diagram. This phase diagram is adapted from \cite{liu2013superconductivity}.
}
\end{figure*}

\end{document}


\title{Supplementary Information for “Chiral Pseudogap Metal Emerging from a Disordered van der Waals Mott Insulator 1$T$-TaS$_{2-x}$Se$_{x}$”}

\author{Hyunjin Jung}
	\affiliation{Center for Artificial Low Dimensional Electronic Systems, Institute for Basic Science (IBS), Pohang 37673, Republic of Korea}
	\affiliation{Department of Physics, Pohang University of Science and Technology, Pohang 37673, Republic of Korea}

\author{Jiwon Jung}
	\affiliation{Center for Artificial Low Dimensional Electronic Systems, Institute for Basic Science (IBS), Pohang 37673, Republic of Korea}
	\affiliation{Department of Physics, Pohang University of Science and Technology, Pohang 37673, Republic of Korea}

\author{ChoongJae Won}
	\affiliation{Center for Artificial Low Dimensional Electronic Systems, Institute for Basic Science (IBS), Pohang 37673, Republic of Korea}
	\affiliation{Laboratory for Pohang Emergent Materials, POSTECH, Pohang 37673, Republic of Korea}
    \affiliation{MPPC-CPM, Max Planck POSTECH/Korea Research Initiative, Pohang 37673, Republic of Korea}

\author{Hae-Ryong Park}
	\affiliation{Center for Artificial Low Dimensional Electronic Systems, Institute for Basic Science (IBS), Pohang 37673, Republic of Korea}
	\affiliation{Department of Physics, Pohang University of Science and Technology, Pohang 37673, Republic of Korea}    

\author{Sang-Wook Cheong}
	\affiliation{Laboratory for Pohang Emergent Materials, POSTECH, Pohang 37673, Republic of Korea}
    \affiliation{MPPC-CPM, Max Planck POSTECH/Korea Research Initiative, Pohang 37673, Republic of Korea}
	\affiliation{Rutgers Center for emergent Materials and Department of Physics and Astronomy, Rutgers University, NJ, USA}    

\author{Jaeyoung Kim}
	\affiliation{Center for Artificial Low Dimensional Electronic Systems, Institute for Basic Science (IBS), Pohang 37673, Republic of Korea}

\author{Gil Young Cho}
	\affiliation{Center for Artificial Low Dimensional Electronic Systems, Institute for Basic Science (IBS), Pohang 37673, Republic of Korea}
	\affiliation{Department of Physics, Pohang University of Science and Technology, Pohang 37673, Republic of Korea}
    \affiliation{Department of Physics, Korea Advanced Institute of Science and Technology, Daejeon 34141, Republic of Korea}
      
\author{Han Woong Yeom}
	\email{yeom@postech.ac.kr}
	\affiliation{Center for Artificial Low Dimensional Electronic Systems, Institute for Basic Science (IBS), Pohang 37673, Republic of Korea}
	\affiliation{Department of Physics, Pohang University of Science and Technology, Pohang 37673, Republic of Korea}

\date{\today}

\maketitle


\newpage
\thispagestyle{empty}        
\mbox{}

\begin{figure*}
\includegraphics{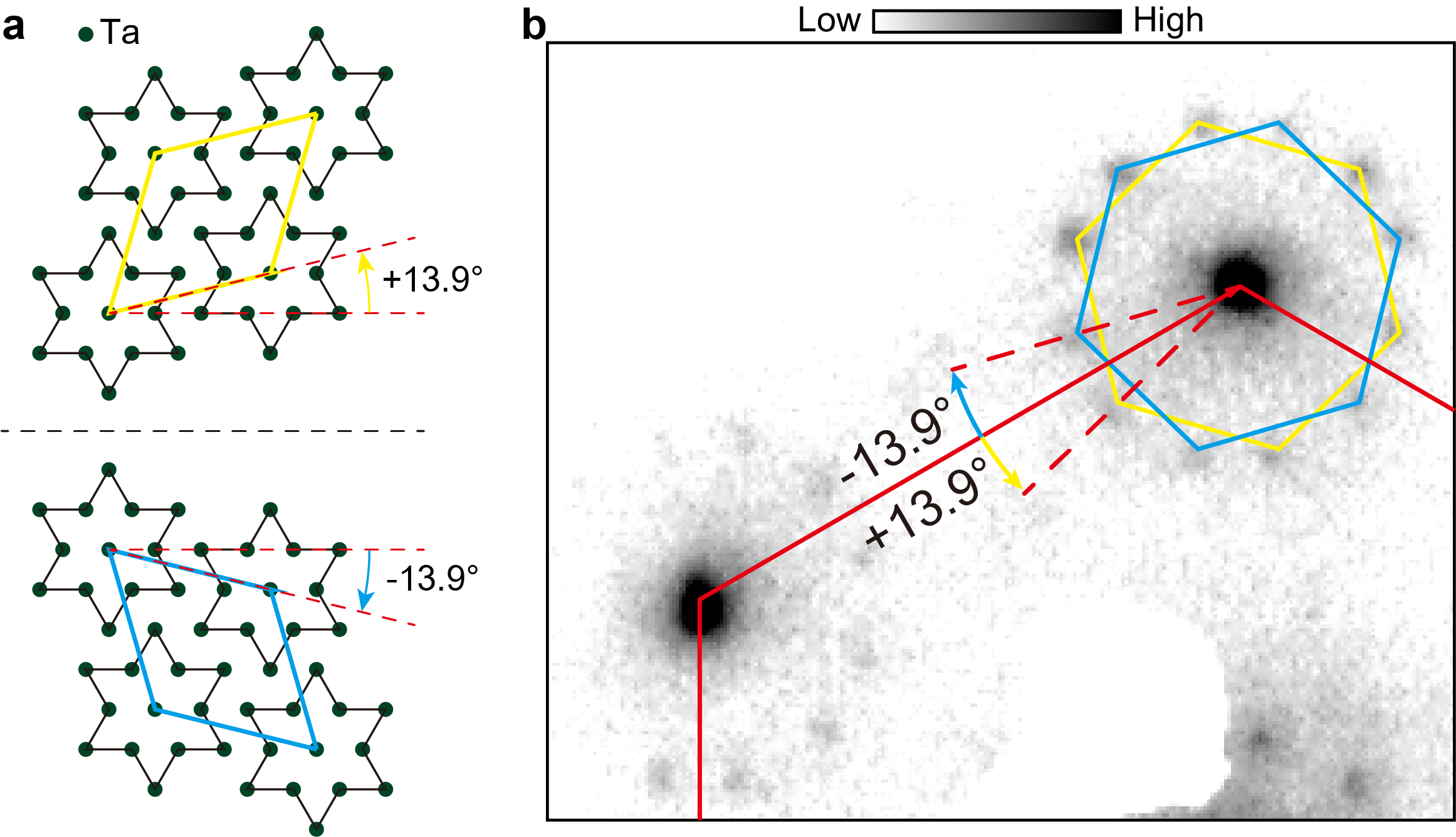}
\caption{\label{Fig.S1}
(a), Schematics of the two ferro-rotational domains in 1T-TaSSe. (b), Grayscale LEED image for 1T-TaSSe (reproduced from Fig. 1e). The yellow and blue guidelines show the diffraction spots generated by two different CWD domains rotated by 27.8 degrees, matched well with FFT image of STM \cite{gao2020pseudogap}.
}
\end{figure*}

\newpage
\thispagestyle{empty}        
\mbox{}

\begin{figure*}
\includegraphics[width=\columnwidth]{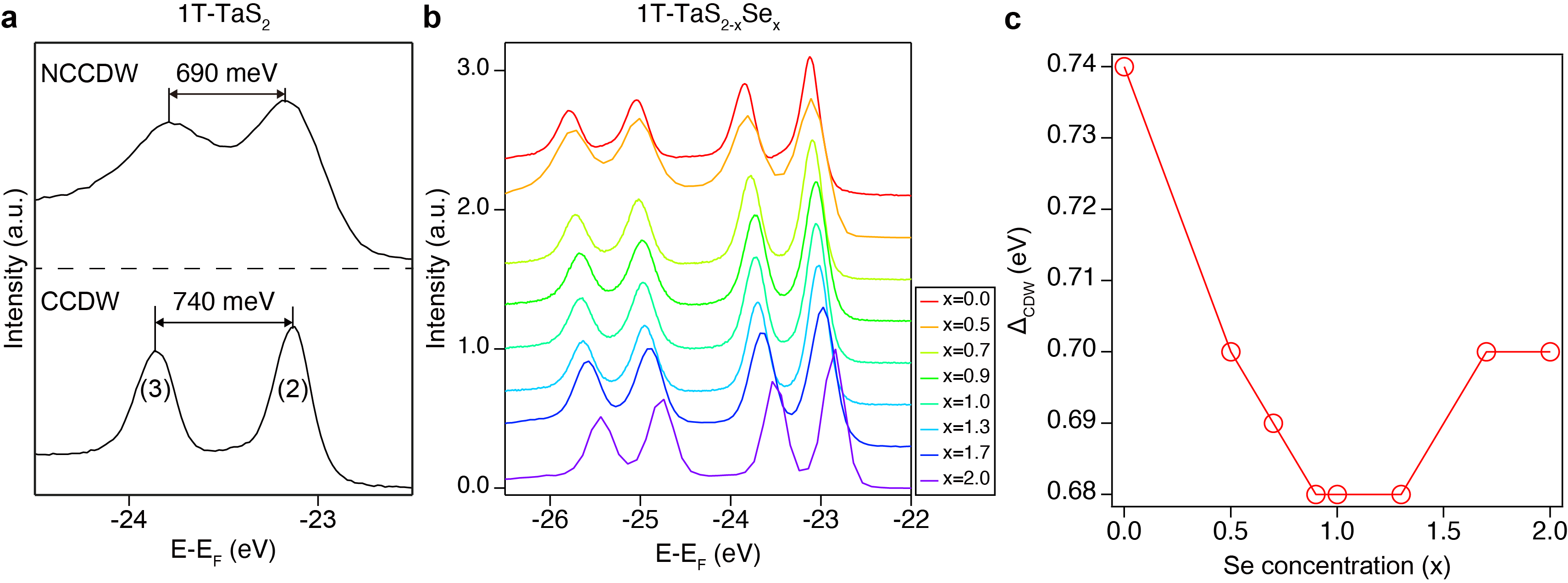}
\caption{\label{Fig.S2}
(a), Ta $4f$ core level spectra of 1$T$-TaS$_{2}$, showing energy splitting of 690 and 740 meV, respectively. Labeled (2) and (3) represent the contribution from the inner and outer Ta atoms within the David star cluster, respectively. (b), Ta $4f$ core level spectra of 1$T$-TaS$_{2-x}$Se$_{x}$. The data for x=0.5 and x=2 were obtained using a photon energy of 200 eV, while the core level data for x=0.0, 0.7, 0.9, 1.0, 1.3, 1.7, and 2.0 were obtained using a photon energy of 65 eV. (c), The plot of $\Delta_{CDW}$ represents the average CDW splitting of the Ta 4$f_{5/2}$ and Ta 4$f_{7/2}$ core level as a function of Se concentration.
}
\end{figure*}

\newpage
\thispagestyle{empty}        
\mbox{}

\begin{figure*}
\includegraphics{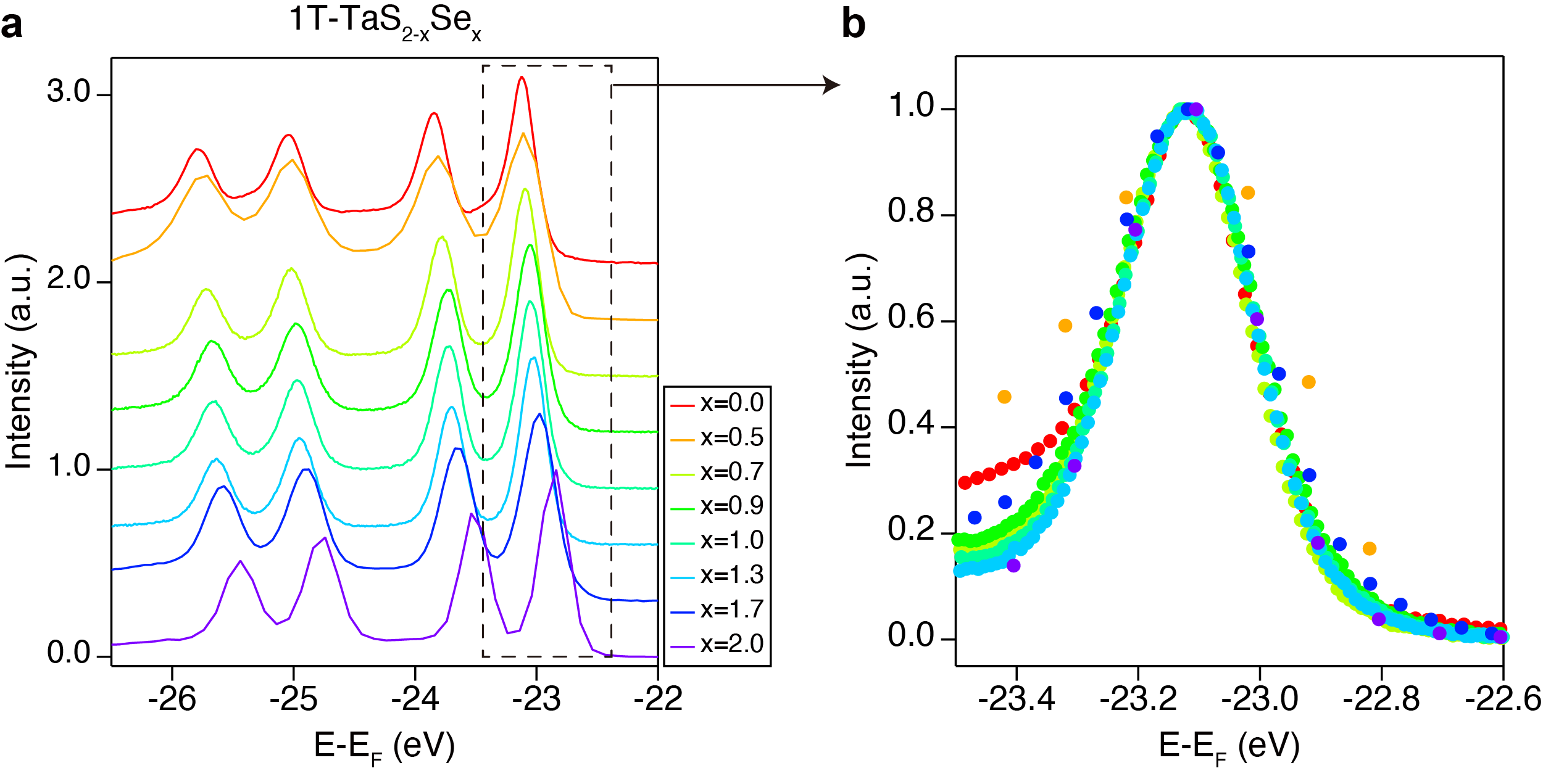}
\caption{\label{Fig.S3}
(a) Ta 4f core-level spectra of 1$T$-TaS$_{2-x}$Se$_{x}$ (reproduced from Fig. S2(b). (b) The same spectra as in (a), horizontally translated so that they overlap at the dashed line. The widths of the Ta 4f core level photoemission spectra are rather consistent for different sample compositions, while that of x=0.5 has a noticeably larger width. We thus interpret that such a width variation is not systematic but due to the different degrees of disorder onto the CDW order itself in the samples. In turn, these data suggest that the CDW order of the samples was maintained rather uniform over different compositions except for that with x=0.5. We used a different photon energy for x=0.5 and x=2.0 (200 eV) from other compositions (65 eV). Thus, the spectra at these two compositions can be broader than the others. However, since the width of the spectra for x=2.0 is consistent with those of others, we conclude the photon energy difference has a little effect on the total spectral broadening.
}
\end{figure*}

\newpage
\thispagestyle{empty}        
\mbox{}

\begin{figure*}
\includegraphics{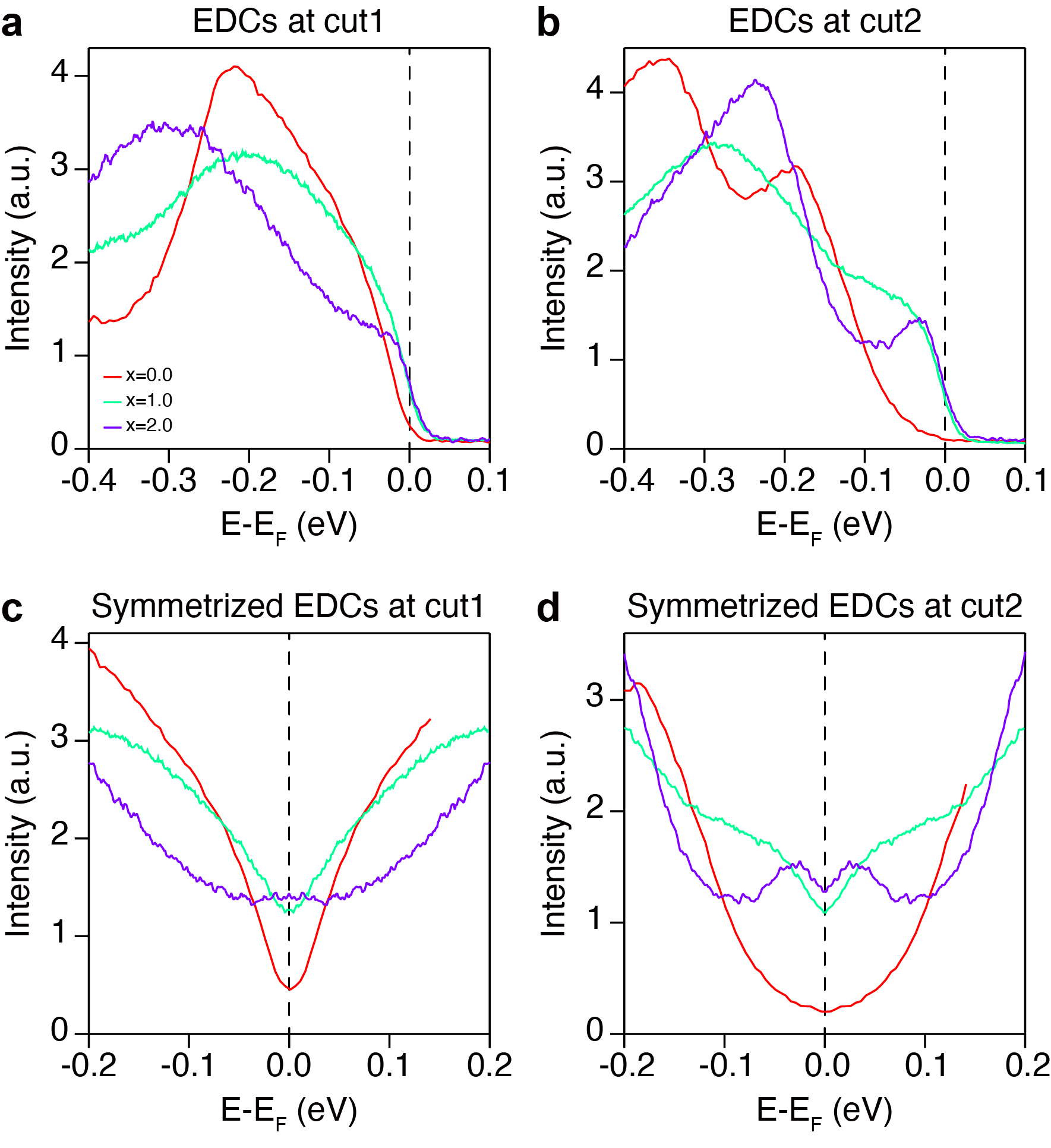}
\caption{\label{Fig.S4}
Direct comparison of energy distribution curves (EDCs) and symmetrized EDCs for x=0, 1.0, and 2.0. (a) and (b) show the normalized EDCs for cut1 and cut2. (c) and (d) present the corresponding symmetrized EDCs.
}
\end{figure*}

\newpage
\thispagestyle{empty}        
\mbox{}

\begin{figure*}
\includegraphics{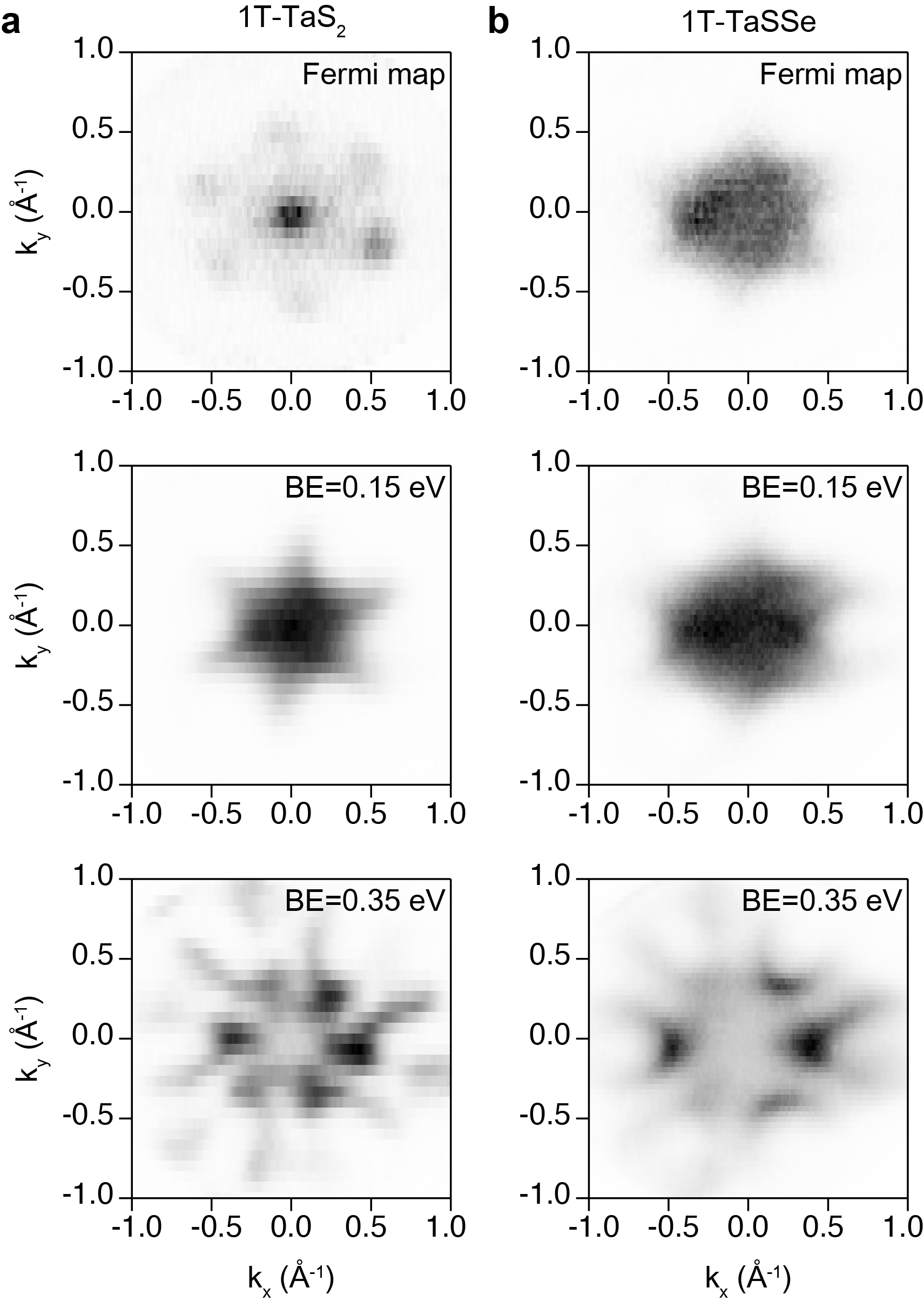}
\caption{\label{Fig.S5}
Constant energy contour maps of (a), 1$T$-TaS$_{2}$ and (b), 1$T$-TaSSe at the binding energies of 0, 0.15 eV, and 0.35 eV.
}
\end{figure*}

\newpage
\thispagestyle{empty}        
\mbox{}

\begin{figure*}
\includegraphics[width=\columnwidth]{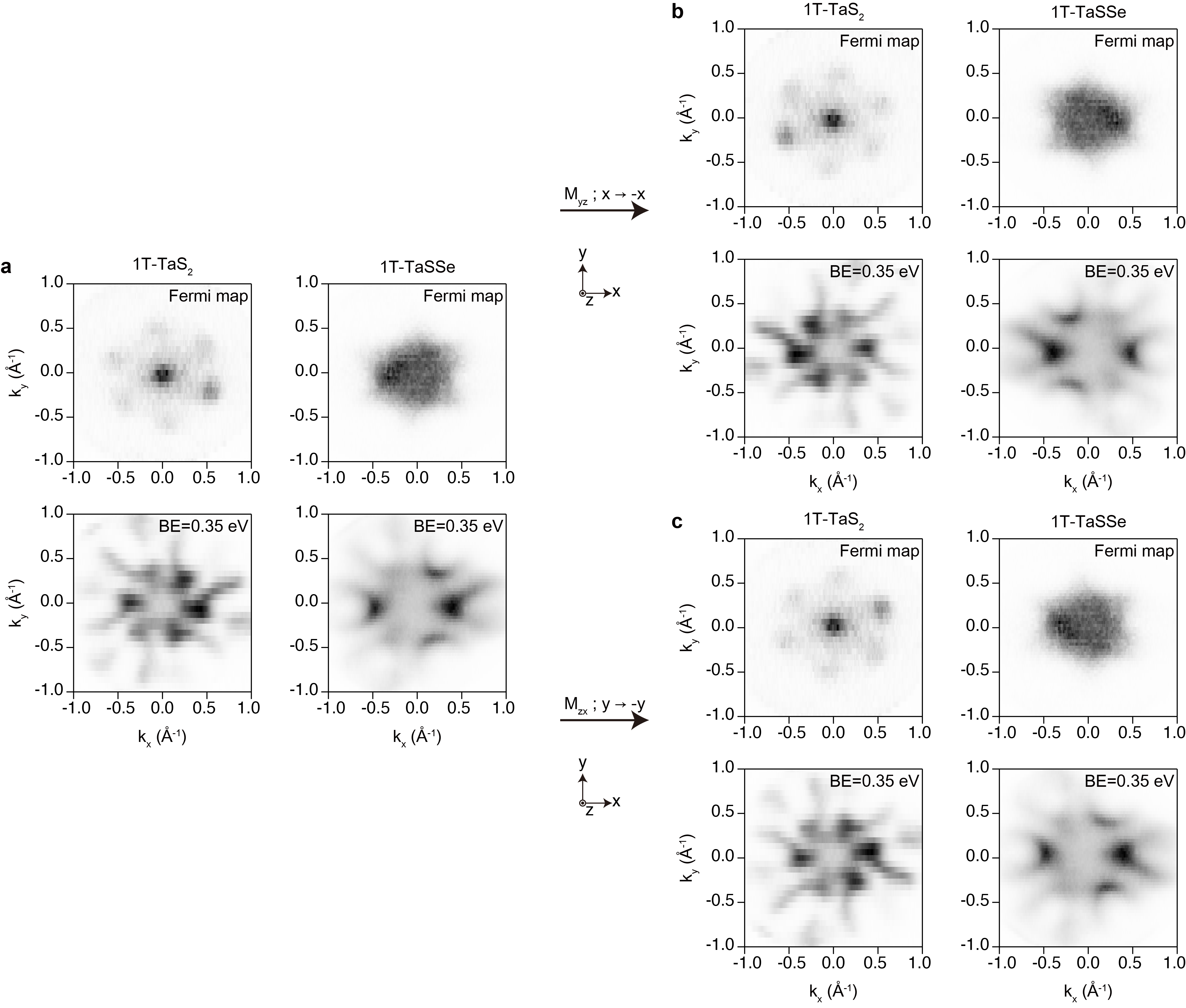}
\caption{\label{Fig.S6}
Constant energy contours of 1$T$-TaS$_{2}$ (left) and 1$T$-TaSSe (right) at Fermi level and BE=0.35eV, respectively. The arrow labeled indicates the mirror-reflection operation along the $k_{x}$ and $k_{y}$ axis. The clear deviation from mirror symmetry in both materials highlights their chiral Fermi map topology.
}
\end{figure*}

\newpage
\thispagestyle{empty}
\mbox{}

\begin{figure*}
\includegraphics{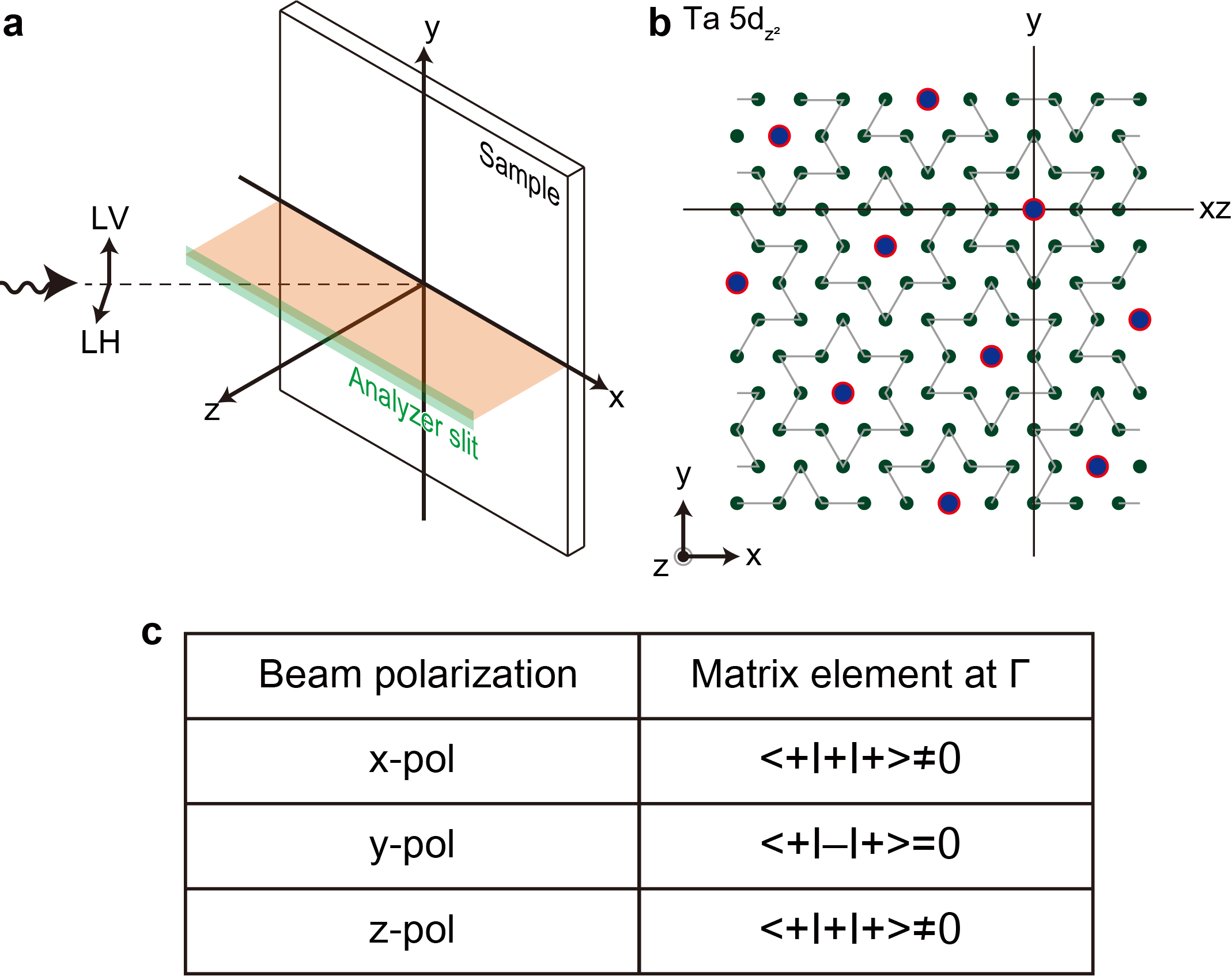}
\caption{
\label{Fig.S7}
Matrix element analysis for Ta $5d_{z^{2}}$ orbitals. (a), Schematics of ARPES setup for the polarization-dependent measurements. (b), illustration of the atomic arrangement for the Ta $5d_{z^{2}}$ orbital. (c), Table of the photoemission selection rule for xz parity at $\overline{\Gamma}$.
}
\end{figure*}

The one-electron dipole matrix element, known as dipole selection rule, can be described as follow \cite{damascelli2003angle, fukutani2021detecting}:
\begin{align}
   M_{fi} & \equiv \left< \phi_{f}|\textbf{\textit{A}} \cdot \textbf{\textit{p}}|\phi_{i} \right>,\\
   & =\left< \phi_{f}|A_{x}p_{x}|\phi_{i} \right>+\left< \phi_{f}|A_{y}p_{y}|\phi_{i} \right>+\left< \phi_{f}|A_{z}p_{z}|\phi_{i} \right>,
\end{align}

where $\phi_{f}$ and $\phi_{f}$ are the final and initial states of electron, \textbf{\textit{A}} is electromagnetic vector potential of photon, and \textbf{\textit{p}} is the electronic momentum operator. In our ARPES setup, the mirror plane is aligned along the xz-plane, resulting in both x- and z-polarization exhibiting even parity, while the y-polarization possesses odd parity. Give that the Ta $5d_{z^{2}}$ orbital is oriented along the z-axis, it maintains even parity with respect to the mirror plane. This symmetry consideration leasds to the matrix element table presented in Fig. S3(c). Specifically, only the y-polarized component carries odd parity, implying that electronic bands contributed from Ta $5d_{z^{2}}$ orbital are detectable exclusively under linear horizontal (LH) polarization. Consequently, the observation of these Ta $5d_{z^{2}}$-dominated bands is restricted to the LH polarization.

\newpage
\thispagestyle{empty}
\mbox{}

\begin{figure*}
\includegraphics{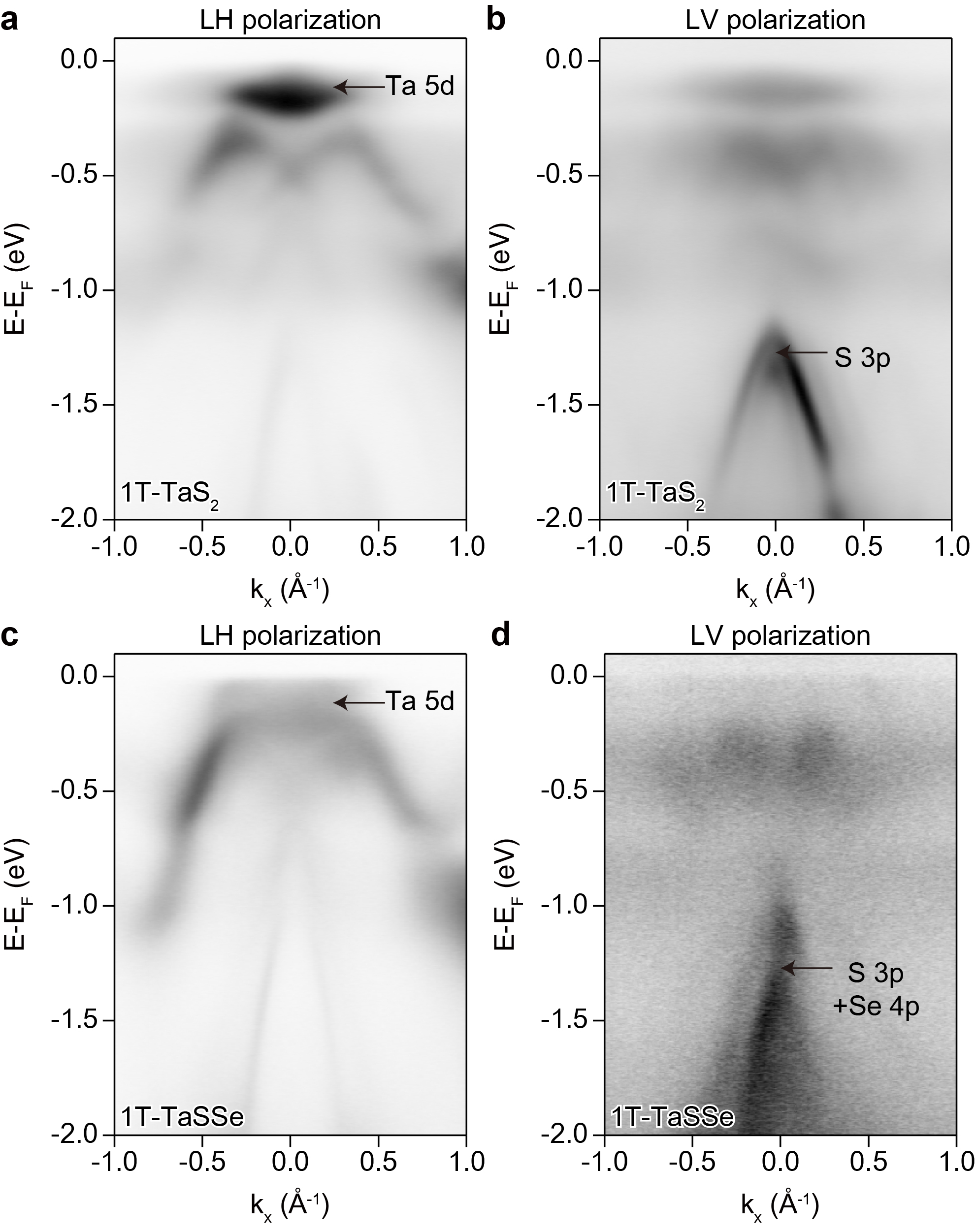}
\caption{
\label{Fig.S8}
Polarization-dependent ARPES spectra of 1$T$-TaS$_{2}$ and 1$T$-TaSSe. (a)-(b), ARPES intensity map with linear horizontal (LH) polarization and linear vertical (LV) polarization of 1$T$-TaS$_{2}$ at 80 K, respectively. (c)-(d), ARPES intensity map with linear horizontal (LH) polarization and linear vertical (LV) polarization of 1$T$-TaSSe at 70 K, respectively. The LH polarization primarily reveals the Ta 5$d$ states, while the LV polarization highlights the contributions from the S 3p+Se 4p states.
}
\end{figure*}

 \newpage
\thispagestyle{empty}
\mbox{}

\begin{figure*}
\includegraphics{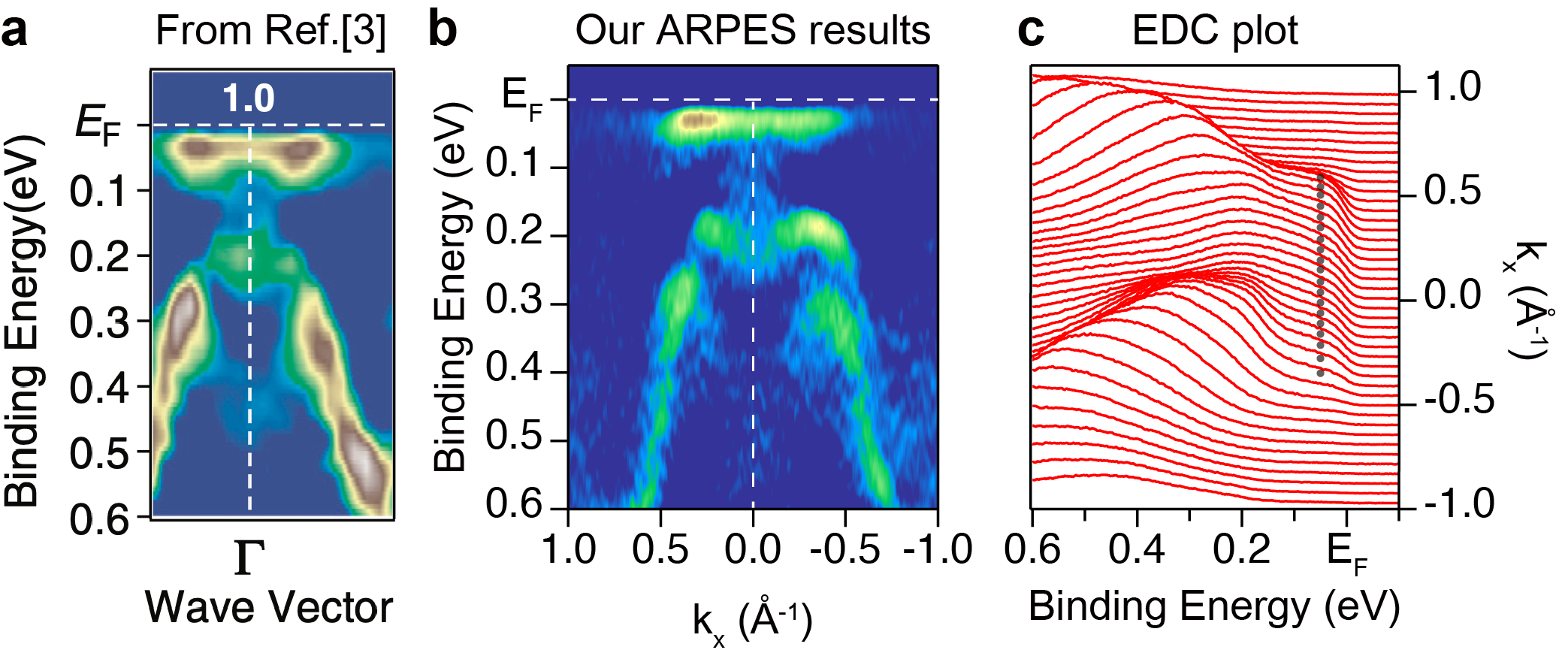}
\caption{
\label{Fig.S9}
Direct comparison of the electronic states of 1$T$-TaSSe between (a) Ref.\cite{ang2013superconductivity} and (b) our second-derivative ARPES results. (c) shows the EDC plot of the raw data from (b). The gray dots indicate the pseudogap edges near the Fermi level ($E_{F}$).
}
\end{figure*}
 
\newpage
\thispagestyle{empty}
\mbox{}

\begin{figure*}
\includegraphics{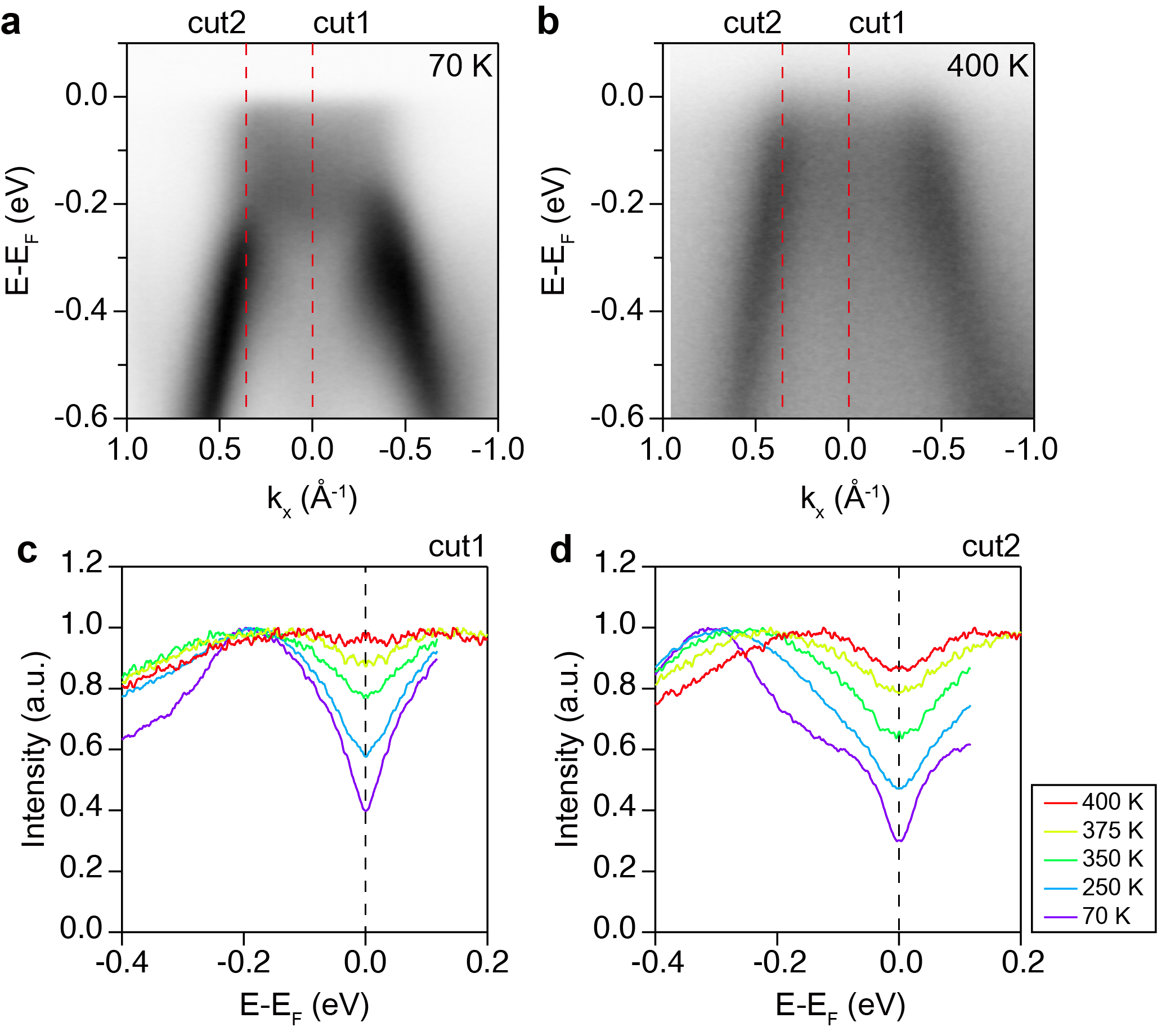}
\caption{
\label{Fig.S10}
Temperature-dependent ARPES measurements of 1$T$-TaSSe. (a-b), Electronic band structure at 70 K and 400 K, respectively. (c-d), Energy distribtuion curves (EDCs) at cut1 and cut2 with increasing temperature, respectively.
}
\end{figure*}

\newpage
\thispagestyle{empty}
\mbox{}

\begin{figure*}
\includegraphics{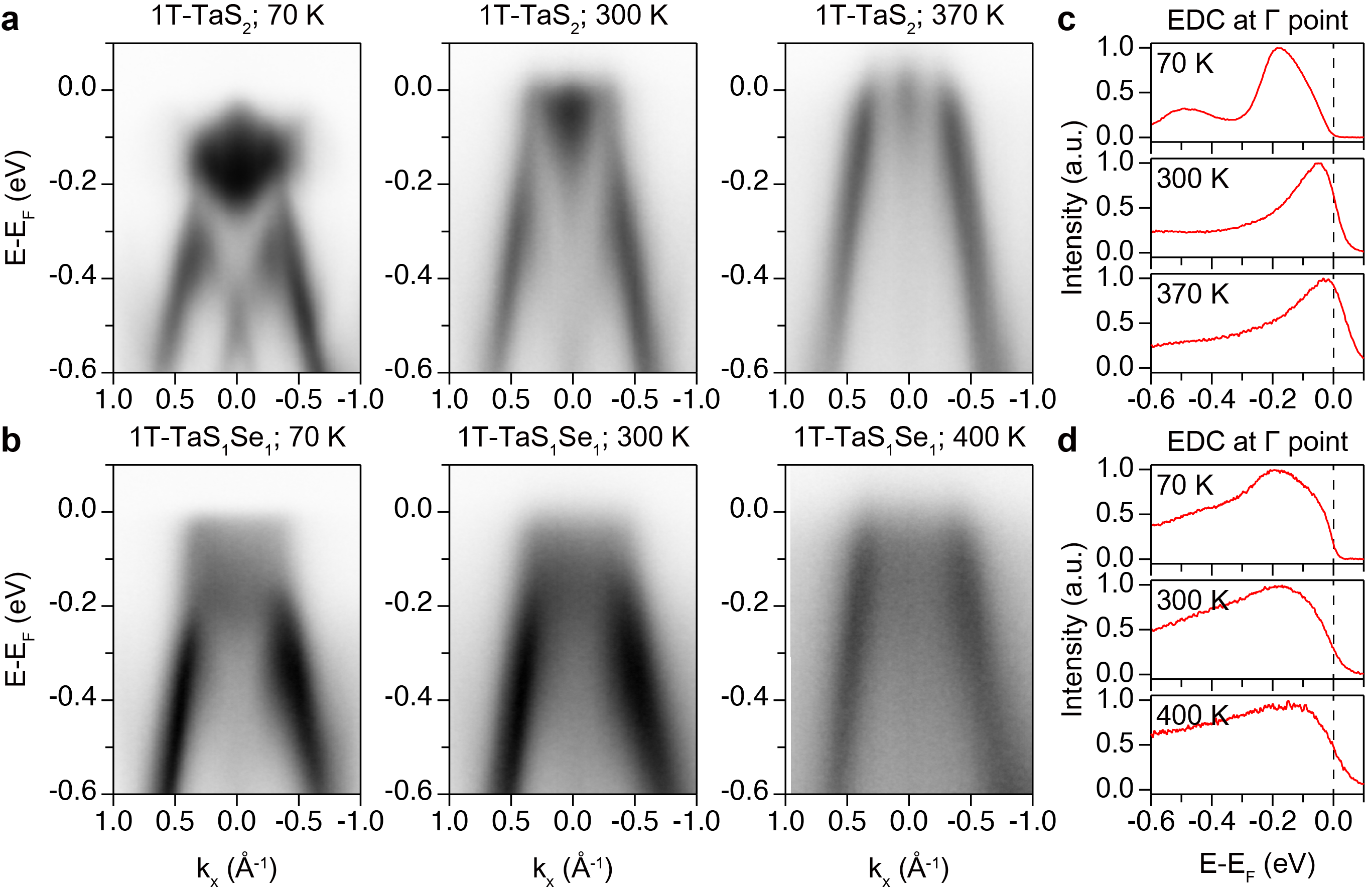}
\caption{
\label{Fig.S11}
Direct comparison of the temperature-dependent ARPES results between pristine 1$T$-TaS$_{2}$ and 1$T$-TaS$_{1}$Se$_{1}$. (a), Electronic band structure of pristine 1$T$-TaS$_{2}$ at 70 K, 300 K, and 370 K. (b), Electronic band structure of pristine 1$T$-TaS$_{1}$Se$_{1}$ at 70 K, 300 K, and 400 K. (c-d), Energy distribution curves (EDCs) at the $\Gamma$ point.
}
\end{figure*}

\newpage
\thispagestyle{empty}
\mbox{}

\begin{figure*}
\includegraphics{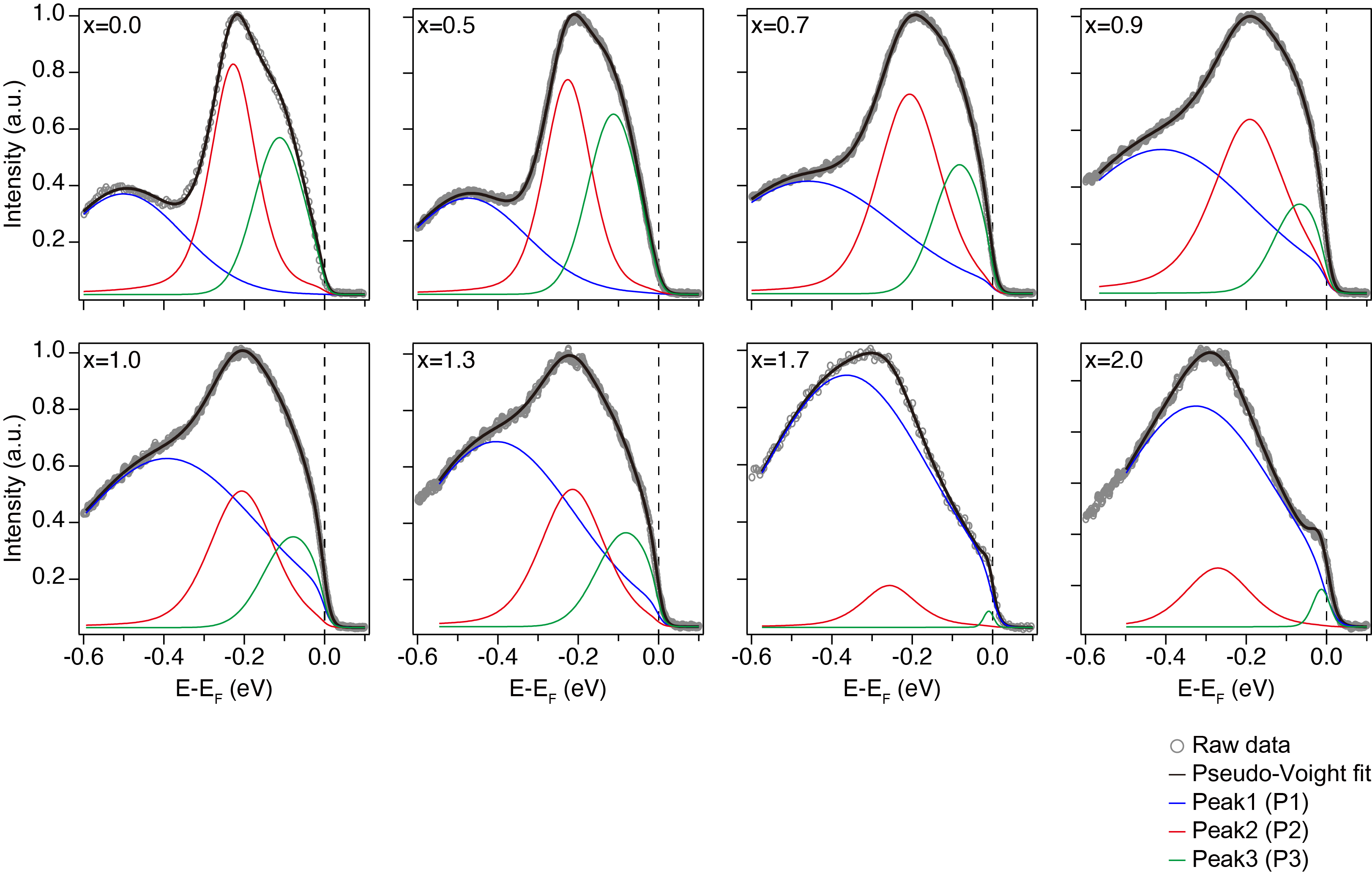}
\caption{
\label{Fig.S12}
Pseudo-Voigt fits of EDC spectra at $\overline{\Gamma}$ point for x=0.0, 0.5, 0.7, 0.9, 1.0, 1.3, 1.7 and 2.0.
}
\end{figure*}

\bibliography{PG_TSS_ref}